\documentclass[aps,pre,onecolumn,english,showpacs,showkeys,preprintnumbers,floatfix,nofootinbib,11pt]{revtex4-2}
\usepackage{eurosym}
\usepackage{amssymb}
\usepackage{graphicx,color}
\usepackage{epsfig}
\usepackage{rotating}
\usepackage[T1]{fontenc}
\usepackage{latexsym,epsfig}
\usepackage[latin9]{inputenc}
\usepackage{amsfonts}
\usepackage{dcolumn}
\usepackage{bm}
\usepackage{babel}
\usepackage{hyperref}
\usepackage{amsmath,mathrsfs}
\usepackage{xcolor}

\newcommand{\Nbin}{N_{\text{bin}}}

\pdfoutput=1

\begin{document}

\title{A spectral investigation of criticality and crossover effects in two and three dimensions: Short timescales with small systems in minute random matrices}
\author{Eliseu Venites Filho $^{1}$, Roberto da Silva$^{1}$, J. R. Drugowich
de Felicio$^{2}$}
\address{1 - Instituto de F{\'i}sica, Universidade Federal do Rio Grande do Sul,
Av. Bento Gon{\c{c}}alves, 9500 - CEP 91501-970, Porto Alegre, Rio Grande do
Sul, Brazil\\ 
2 - Departamento de F\'{\i}sica, Faculdade de Filosofia Ci\^{e}ncias e Letras de
Ribeir\~{a}o Preto, Universidade de S\~{a}o Paulo, 
Av. dos Bandeirantes 3900, Ribeir\~{a}o Preto, S\~{a}o Paulo, Brazil}

\begin{abstract}
Random matrix theory, particularly using matrices akin to the Wishart
ensemble, has proven successful in elucidating the thermodynamic
characteristics of critical behavior in spin systems across varying
interaction ranges. This paper explores the applicability of such methods in
investigating critical phenomena and the crossover to tricritical points
within the Blume-Capel model. Through an analysis of eigenvalue mean,
dispersion, and extrema statistics, we demonstrate the efficacy of these
spectral techniques in characterizing critical points in both two and three
dimensions. Crucially, we propose a significant modification to this
spectral approach, which emerges as a versatile tool for studying critical
phenomena. Unlike traditional methods that eschew diagonalization, our
method excels in handling short timescales and small system sizes, widening
the scope of inquiry into critical behavior.
\end{abstract}

\maketitle

\section{Introduction}

The phenomenology of critical phenomena, encompassing phase transitions in
diverse contexts, stands as a cornerstone within the framework of
Statistical Mechanics theory. Initially conceived within the realm of
many-body physics, it has evolved into a concept with far-reaching
applications spanning disciplines such as economics, network theory,
sociophysics, game theory, and numerous others \cite{Stanley, Bouchaud, Castellano, Szabo, Barabasi}.

In spin systems, a particularly effective approach to delve into critical
phenomena involves conducting short-time dynamics studies. This method
entails preparing systems with carefully chosen initial conditions and then
analyzing temporal averages. Through this analysis, the manifestation of
critical behavior in the system is revealed via power-law dynamics \cite%
{Janssen,Janssen2,Henkel,Zhengprimordial,Huse,Albano}.

Such an endeavor can be pursued in models exhibiting up-down symmetry \cite%
{Grinstein,TomePRE1998,TaniaMario2014}, as well as in models featuring
absorbing states within the universality class of direct percolation \cite%
{Henkel,Janssen3,Grassberger,Dickman,Hinrichsen2000,SilvaDickman}.
Additionally, these studies extend to tricritical points (TCP), encompassing
two-dimensional and three-dimensional short-range systems, as well as
systems under mean-field approximation \cite%
{Janssen1994,SilvaBC3D2022,SilvaBJPMF,Silva2002,Silva2013}, each with their
distinctive characteristics and intricacies.

Utilizing time-dependent simulations to investigate these techniques
presupposes the capability to compress temporal evolution while contending
with inherently large systems, a challenge frequently encountered in
equilibrium simulations. Moreover, equilibrium simulations grapple with the
issue of critical slowing down. Nowadays simulations of two-dimensional
systems typically encompass linear dimensions denoted by $L$, conservatively
ranging in the order of hundreds or, with more generous allocations,
extending into the thousands. These simulations leverage parallelization
techniques and GPU applications to effectively handle computational demands.

This suggests that exploring non-conventional methods is intriguing. Upon delving into the literature, the concept of random matrices reveals fascinating aspects. Originating from the intricate description of the distribution of energy levels in heavy nuclei proposed by E. Wigner \cite{Wigner,Wignerb,Wigner2,Mehta}, its connection with the thermodynamics of Coulomb gases primarily stems from the works of Dyson \cite{Dyson,Dyson2,Dyson3}.  

Alternatively, utilizing individual time series to construct appropriate
matrices, known as Wishart matrices \cite{Wishart,Livan}, offers a particularly
intriguing avenue for delving into the statistical mechanics of spin systems
and their idiosyncrasies. This approach is underscored by recent
contributions \cite{Vinayak2014,Biswas2017,RMT2023,RMT-2}, as these matrices
encapsulate the time-correlations of specific random variables described by
stochastic processes, which can be modeled by Langevin equations or
simulated via Monte Carlo (MC) Markov chains. It's noteworthy to mention that
while the integrability aspects of lattice spin systems have been explored
using random matrices, the emphasis has traditionally been on direct
examination of the Hamiltonians of such systems \cite{RMTMaillard}, rather
than on correlation matrices as in the previously mentioned works.

In contrast to other methods found in the literature, the approach advocated
by one of the authors of this paper in \cite{RMT2023} diverges significantly
by eschewing the use of matrices whose dimensions correspond directly to the
number of lattice sites. Such an approach, while theoretically sound, can be
computationally prohibitive due to the sheer scale of the matrices involved.
Instead, the proposed method operates effectively with dimensions on the
order of a few hundred, derived from various short-term evolution samples of the
magnetization system. This strategy, as evidenced in our prior research \cite%
{RMT2023,RMT-2}, proves to be both computationally efficient and
theoretically robust.

In this study, we aim to extend the applicability of random matrix
methodology by investigating models that exhibit tricritical points, with a
particular emphasis on the transition from critical points to the tricritical
one. Through careful analysis of the spectra derived from Wishart-like
matrices, which capture the correlations among diverse magnetization time
series, we showcase the effectiveness of our approach in elucidating the
phase diagram of the Blume-Capel (BC) model in both two and three
dimensions. Throughout this paper, we present our findings, highlighting how
this methodology adeptly reveals the intricate characteristics of complex
systems like the BC model.

Furthermore, we demonstrate that this method, despite its efficiency and
suitability for short times, operates effectively for small systems,
contrastively to the standard MC method.

In the next section, we introduce the Wishart-like method alongside a
concise overview of the established properties of the BC model. It's crucial
to emphasize that our primary focus lies not on the intricacies of the model
itself. Rather, our goal was to select the simplest model featuring a
tricritical point in both two and three dimensions to validate our study.

In the following section (Section \ref{Sec:Results}), we unveil our
principal findings, demonstrating the efficacy of our method in accurately
pinpointing the critical points of both the two-dimensional and
three-dimensional versions of the model. Additionally, we illustrate how our
approach adeptly captures the crossover effects in a spectral manner.
Furthermore, a novel aspect explored in this study, not previously
investigated in our prior works, is the successful application of extreme value
statistics in establishing the criticality of such systems. 

Lastly, we demonstrate the effectiveness of our method even for very small
systems. We illustrate that although there may be a computational cost
associated with diagonalizing matrices, this cost is offset by the
advantages gained from shorter computation times and smaller system sizes.
The paper concludes with a summary of key findings and conclusions.

\section{Random Matrices, Critical, and Tricritical Points in the
Blume-Capel Model}

\label{Sec:Model}

Tricritical points play a vital role in contemporary research within
condensed matter theory and statistical mechanics. The pioneering discovery
of the first tricritical point in He-3 and He-4 mixtures by R. B. Griffiths
in 1970 marked a significant milestone \cite{Griffiths}. Subsequently,
Griffiths, along with Blume and Emery, introduced the BEG
(Blume-Emery-Griffiths) model in 1971, which provided a framework to
replicate the thermodynamic behavior observed in these mixtures \cite{Blume1971}.
This model, based on a spin-1 Ising model, has since become a
cornerstone in the study of tricritical phenomena.

In its broader scope, the Hamiltonian can be expressed as:%
\begin{equation*}
\mathcal{H}=-K\sum_{\left\langle ij\right\rangle
}s_{i}^{2}s_{j}^{2}-H_{3}\sum_{\left\langle ij\right\rangle
}s_{i}s_{j}(s_{i}+s_{j})-J\sum_{\left\langle ij\right\rangle
}s_{i}s_{j}+D\sum_{i=1}^{N}s_{i}^{2}-H\sum_{i=1}^{N}s_{i}
\end{equation*}

The initial term delineates the crystalline interaction among the spins,
whereas the subsequent term denotes the multispin interaction among them.
Finally, the third, fourth, and last terms represent the Ising interaction
between spin pairs, anisotropic interaction, and the interaction of spins
with an external magnetic field, respectively.

Indeed, an even more straightforward variation of this model, commonly
referred to as the Blume-Capel (BC) model \cite{Blume,Capel}, is delineated by the
Hamiltonian:%
\begin{equation}
\mathcal{H}=-J\sum_{\langle i,j\rangle}s_{i}s_{j}+D\sum_{i}s_{i}^{2}-H\sum_{i}s_{i}
\end{equation}

where each spins can hold the values $s_{i}\in \{-1,0,+1\}$, it would be
enough to present the existence of a tricritical point in both two and
three dimensions, separating a critical line of the first order line.

The first term models the local interaction between the spins, with $J>0$
representing the interaction strength and $\langle i,j\rangle $ indicating
that the interaction occurs between nearest neighbor pairs of sites $i$ and
$j$. The parameter $D$ is called the anisotropy field and is responsible for
zero-field splitting, resulting in an increase in energy for $s_{i}=\pm 1 $
states even in the absence of an external magnetic field. Finally, the third
term models the interaction of the system with an external magnetic field of
intensity $H$, which we will assume is not present.

\begin{figure*}[tbp]
\begin{center}
\includegraphics[width=1.0\columnwidth]{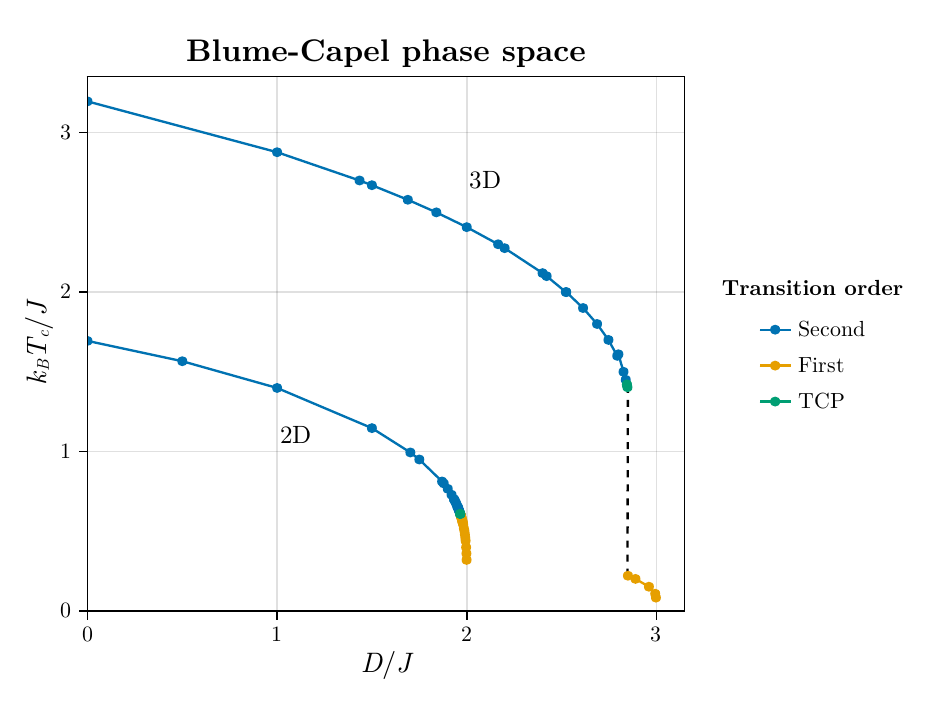}
\end{center}
\caption{The phase diagrams for the two and three-dimensional BC
models are depicted. The points utilized in our numerical experiments are
extracted from Butera and Pernici \protect\cite{Butera}, serving as
foundational data for the investigations conducted in this study. }
\label{Fig:Phase_diagram}
\end{figure*}

In this scenario, the model delineates a critical line (CL) culminating in a
tricritical point (TCP). Subsequently, it exhibits a first-order line (FOL),
as illustrated in Figure \ref{Fig:Phase_diagram} for both the two-dimensional
and three-dimensional versions of the BC model in the absence of an external magnetic field.

This figure is utilized only for pedagogical reasons in this work since all
points used in this current work uses the points estimated in this line that
were very didactically obtained by Butera and Pernici \cite{Butera}. Sure, a
lot of estimates shown in such reference were re-obtained by the authors
since there was a evolution of estimates with very different methods along
of many years that include the a lot of good works (see for example \cite%
{Beale,Deserno,Plascak,Hasenbusch}). Thus the question is if we can detect
the critical line of the BC model for different values of $D$ using
Wishart-random matrices spectra and how the method responds to the crossover
between CL and FOL intermediated by the TCP point.

Therefore, leveraging a framework established in prior research, this study
demonstrates our capability to localize critical points within two and
three-dimensional BC models, while also investigating the existing
crossover phenomena. Subsequently, the following section provides a concise
overview of the random matrix methodology employed in our approach, along
with the quantities slated for estimation within this method.

\subsection{Wishart-like matrices and spin systems}

In our analysis, we introduce the magnetization matrix element $m_{tj}$,
which denotes the magnetization of the $j$th time series at the $t$th Monte
Carlo (MC) step in a system comprising $N=L^{d}$ spins. For simplicity, we
set $d=2$, as it represents the minimal dimension for the manifestation of
phase transitions in short-range interaction systems. In this context, $t$
ranges from 1 to $N_{MC}$, and $j$ ranges from 1 to $N_{sample}$, thereby
constructing the magnetization matrix $M$ with dimensions $N_{MC} \times N_{sample}$.

To delve into spectral properties, an intriguing approach is to shift our
focus away from $M$ and instead examine the square matrix of size $N_{sample}\times N_{sample}$:%
\begin{equation*}
G=\frac{1}{N_{MC}}M^{T}M\ ,
\end{equation*}
where each element $G_{ij}$ of $G$ is defined as $G_{ij}=\frac{1}{N_{MC}}%
\sum_{t=1}^{N_{MC}}m_{ti}m_{tj}$, referred to as the Wishart matrix \cite%
{Wishart}. To simplify computations, it is advantageous to transform the
components of the matrix $M$ using the transformed matrix $M^*$, whose
elements are expressed in terms of standard variables as follows:%
\begin{equation*}
m_{tj}^{\ast }=\frac{m_{tj}-\left\langle m_{j}\right\rangle _{t}}{\sqrt{%
\left\langle m_{j}^{2}\right\rangle _{t}-\left\langle m_{j}\right\rangle
_{t}^{2}}},
\end{equation*}
where $\left\langle m_{j}^{k}\right\rangle _{t}=\frac{1}{N_{MC}}%
\sum_{i=1}^{N_{MC}}m_{ij}^{k}\ $. This transformation facilitates subsequent
analysis and calculations.

Thereby:%
\begin{equation}
G_{ij}^{\ast }=\frac{\left\langle m_{i}m_{j}\right\rangle -\left\langle
m_{i}\right\rangle \left\langle m_{j}\right\rangle }{\sigma _{i}\sigma _{j}}
\label{Eq:Correlation}
\end{equation}

where $\left\langle m_{i}m_{j}\right\rangle _{t}=\frac{1}{N_{MC}}%
\sum_{t=1}^{N_{MC}}m_{ti}m_{tj}$ and $\sigma _{i}=\sqrt{\left\langle
m_{i}^{2}\right\rangle -\left\langle m_{i}\right\rangle ^{2}}$.
Here it is crucial to expound upon a pivotal calculation that elucidates the
application of these matrices in greater detail. We consider two distinct
time evolution samples of the magnetization per spin denoted as $m_{ti}$ and $m_{tj}$,
where $t=1,...,N_{MC}$. In this context:%
\begin{equation*}
m_{tj}=\frac{1}{N}\sum_{k=1}^{N}\sigma _{t,j,k}\text{,}
\end{equation*}

where $\sigma _{t,j,k}$ denotes the value of the $k$-th spin in the $j$-th
evolution or run at time $t$.

We can establish the correlation between these two time series using the
following definition:%
\begin{equation*}
\begin{array}{lll}
\left\langle m_{i}m_{j}\right\rangle _{t} & = & \frac{1}{N_{MC}}%
\sum_{t=1}^{N_{MC}}m_{ti}m_{tj} \\ 
&  &  \\ 
& = & \frac{1}{N^{2}N_{MC}}\sum_{t=1}^{N_{MC}}\left( \sum_{k=1}^{N}\sigma
_{t,i,k}\right) \left( \sum_{l=1}^{N}\sigma _{t,j,l}\right) \\ 
&  &  \\ 
& = & \frac{1}{N^{2}N_{MC}}\sum_{t=1}^{N_{MC}}\left( \sum_{k=1}^{N}\sigma
_{t,i,k}\sigma _{t,j,k}+\sum_{k\neq l=1}^{N}\sigma _{t,i,k}\sigma
_{t,j,l}\right) \\ 
&  &  \\ 
& = & \frac{1}{N^{2}}\sum_{k=1}^{N}\left[ \left( \frac{1}{N_{MC}}%
\sum_{t=1}^{N_{MC}}\sigma _{t,i,k}\sigma _{t,j,k}\right) +\sum_{k\neq
l=1}^{N}\left( \frac{1}{N_{MC}}\sum_{t=1}^{N_{MC}}\sigma _{t,i,k}\sigma
_{t,j,l}\right) \right] \\ 
&  &  \\ 
& = & \frac{1}{N^{2}}\sum_{k=1}^{N}\left\langle \sigma _{i,k}\sigma
_{j,k}\right\rangle _{t}+\frac{1}{N^{2}}\sum_{k\neq l=1}^{N}\left\langle
\sigma _{i,k}\sigma _{j,l}\right\rangle _{t}%
\end{array}%
\end{equation*}

Given that $\sum_{k=1}^{N}\left\langle \sigma _{i,k}\sigma
_{j,k}\right\rangle _{t}=O(N)$, and $\sum_{k\neq l=1}^{N}\left\langle \sigma
_{i,k}\sigma _{j,l}\right\rangle _{t}=O(N^{2})$, it follows that
thermodynamic limit ($N\rightarrow \infty $):%
\begin{equation*}
\left\langle m_{i}m_{i}\right\rangle _{t}\approx \left\langle \frac{1}{N^{2}}%
\sum_{k\neq l=1}^{N}\sigma _{i,k}\sigma _{j,l}\right\rangle _{t} = \frac{1}{%
N^{2}}\left\langle \sigma _{i}\otimes \sigma _{j}\right\rangle _{t}\text{.}
\end{equation*}

When $T>T_{C}$, $\left\langle m_{i}\right\rangle _{t}\approx 0$. This leads
to: $\left\langle m_{i}m_{j}\right\rangle _{t}-\left\langle
m_{i}\right\rangle _{t}\left\langle m_{j}\right\rangle _{t}\approx
\left\langle m_{i}m_{j}\right\rangle _{t}=\frac{1}{N^{2}}\left\langle \sigma
_{i}\otimes \sigma _{j}\right\rangle _{t}$, and we can express the
correlation coefficient (our matrix element of $G$) as:%
\begin{equation*}
\begin{array}{lll}
G_{ij}^{\ast } & \approx & \frac{\left\langle m_{i}m_{j}\right\rangle _{t}}{%
\sqrt{\left\langle m_{i}^{2}\right\rangle _{t}}\sqrt{\left\langle
m_{j}^{2}\right\rangle _{t}}} \\ 
&  &  \\ 
& = & \frac{\left\langle m_{i}m_{j}\right\rangle _{t}}{\left\langle
m_{i}^{2}\right\rangle _{t}} \\ 
&  &  \\ 
& = & \frac{\left\langle \sigma _{i}\otimes \sigma _{j}\right\rangle _{t}}{%
\left\langle \sigma _{i}\otimes \sigma _{i}\right\rangle _{t}}%
\end{array}%
\end{equation*}%
where $\sigma _{i}\equiv (\sigma _{i,1},...,\sigma _{i,N})$\ and $\sigma
_{j}\equiv (\sigma _{j,1},...,\sigma _{j,N})$. Thus $g_{ij}$\ for $T>T_{C}$
is determined by: 
\begin{equation*}
G_{ij}^{\ast }=\frac{\left\langle \sigma _{i}\otimes \sigma
_{j}\right\rangle _{t}}{\left\langle \sigma _{i}\otimes \sigma
_{i}\right\rangle _{t}}
\end{equation*}

This metric assesses the relationship between the temporal averages of
spatial correlations within both inter and intra-time series. By analyzing
both spatial and temporal dimensions, it provides a compelling approach to
delve into spin systems.

Thinking in the general case, when the variables $m_{ij}^*$ are uncorrelated random variables, momentarily forgetting the context of these variables represent magnetization of spin systems, the eigenvalue density $\rho (\lambda )$ of the matrix $G^{\ast }=\frac{1}{N_{MC}} M^{\ast T}M^{\ast }$ conforms to the well-known Marchenko-Pastur (MP) distribution \cite{Marcenko}. For our specific case, we express this distribution as:

\begin{equation}
\rho (\lambda )=\left\{ 
\begin{array}{l}
\dfrac{N_{MC}}{2\pi N_{sample}}\dfrac{\sqrt{(\lambda -\lambda _{-})(\lambda
_{+}-\lambda )}}{\lambda }\ \text{if\ }\lambda _{-}\leq \lambda \leq \lambda
_{+} \\ 
\\ 
0\ \ \ \ \text{otherwise}%
\end{array}%
\right.  \label{Eq:MP2}
\end{equation}%
where $\lambda _{\pm }=1+\frac{N_{sample}}{N_{MC}}\pm 2\sqrt{\frac{N_{sample}%
}{N_{MC}}}.$

Undoubtedly, we expect that for $T \gg T_{c}$, the density of eigenvalues $%
\rho^{\exp}(\lambda)$ obtained from computational simulations approaches $%
\rho(\lambda)$ in Equation \ref{Eq:MP2}, but our method may not necessarily fit
such a distribution perfectly due to residual autocorrelation. The interesting
question is what happens when $T \approx T_{C}$. Moreover, we will utilize
the density $\rho^{\exp}(\lambda)$, obtained from computer simulations, to
determine the critical parameters of spin models.

The moments of $\rho ^{\exp }(\lambda )$ are calculated as: 
\begin{equation}
\left\langle \lambda ^{k}\right\rangle =\frac{\sum_{i=1}^{\Nbin}\lambda
_{i}^{k}\rho ^{\exp }(\lambda _{i})}{\sum_{i=1}^{\Nbin}\rho ^{\exp
}(\lambda _{i})}\text{,}  \label{Eq:Numerical_moments}
\end{equation}%
where $\Nbin$ is the number of bins of the histogram of $%
\rho^{\exp}(\lambda)$. Thus, for $T \gg T_{c}$, we also expect $\overline{%
\lambda ^{k}}$ to approach: 
\begin{equation*}
\begin{array}{lll}
E\left[ \lambda ^{k}\right] & = & \int_{-\infty }^{\infty }\lambda ^{k}\rho
(\lambda )d\lambda \\ 
&  &  \\ 
& = & \dfrac{N_{MC}}{2\pi N_{sample}}\int_{\lambda _{-}}^{\lambda
_{+}}\lambda ^{k-1}\sqrt{(\lambda -\lambda _{-})(\lambda _{+}-\lambda )}%
d\lambda \\ 
&  &  \\ 
& = & \sum\limits_{j=0}^{k-1}\frac{\left( \frac{N_{sample}}{N_{MC}}\right)
^{j}}{j+1}\binom{k}{j}\binom{k-1}{j}%
\end{array}%
\end{equation*}

Explicitly, $E[\lambda] = 1$ and $E[\lambda^2] = \sum\limits_{j=0}^{1}\frac{%
\left(\frac{N_{sample}}{N_{MC}}\right)^j}{j+1}\binom{2}{j}\binom{1}{j} = 1 + 
\frac{N_{sample}}{N_{MC}}$. However, beyond these limits, the behavior of $%
\left\langle \lambda^k\right\rangle$ can provide thermodynamic information
about spin models, as suggested by our previous works \cite{RMT2023,RMT-2}.
In those works, we observed that monitoring $\left\langle \lambda
\right\rangle$ and $\left\langle \left(\Delta \lambda \right)^2\right\rangle
= \left\langle \left(\lambda - \left\langle \lambda \right\rangle
\right)^2\right\rangle$ as a function of $\frac{T}{T_{C}}$ indicates a
minimum of $\left\langle \lambda \right\rangle$ and an inflection point for $%
\left\langle \left(\lambda - \left\langle \lambda \right\rangle
\right)^2\right\rangle$ (or divergence of its derivative) occurs at $T=T_{C}$.

Here, we will demonstrate that this method works effectively for the BC
model in both two and three dimensions, particularly in identifying critical
points and examining its response to the crossover phenomena between CL and
FOL.

\section{Results}

\label{Sec:Results}

Now, we will present our main results. In the first subsection, we showcase
the outcomes of our spectral method concerning the critical points of the BC
model in both two and three dimensions. Following this, in the second
subsection, we extend our investigation to demonstrate the crossover effects
in the model, as captured by the density of maximal eigenvalues of Wishart
matrices.

For our analysis, we construct $N_{run}=1000$ distinct matrices $G^{\ast }$
of size $N_{sample}\times N_{sample}$ for each fixed temperature. Each
matrix is derived from $N_{sample}=100$ magnetization time series, each
comprising $N_{MC}=300$ Monte Carlo steps. These time series are
obtained via MC simulations employing heatbath single spin flip dynamics
for the BC model, resulting in a total of $10^{5}$ eigenvalues used to
construct the histogram for each temperature. All eigenvalues are
categorized into $N_{b}=100$ bins. In the two-dimensional systems, we
utilize a linear dimension of $L=100$, while in three dimensions, we employ $%
L=22$.

An essential aspect for the accurate numerical application of the method
involves utilizing the histogram to compute the eigenvalue moments through
numerical experiments, as per Equation \ref{Eq:Numerical_moments}, and directly
calculating the numerical moments. It is crucial to emphasize this point for
readers intending to apply the method, as we have confirmed that computing
the averages directly does not yield the expected results presented here.

\subsection{Critical points}

\begin{figure*}[tbp]
\begin{center}
\includegraphics[width=1.0%
\columnwidth]{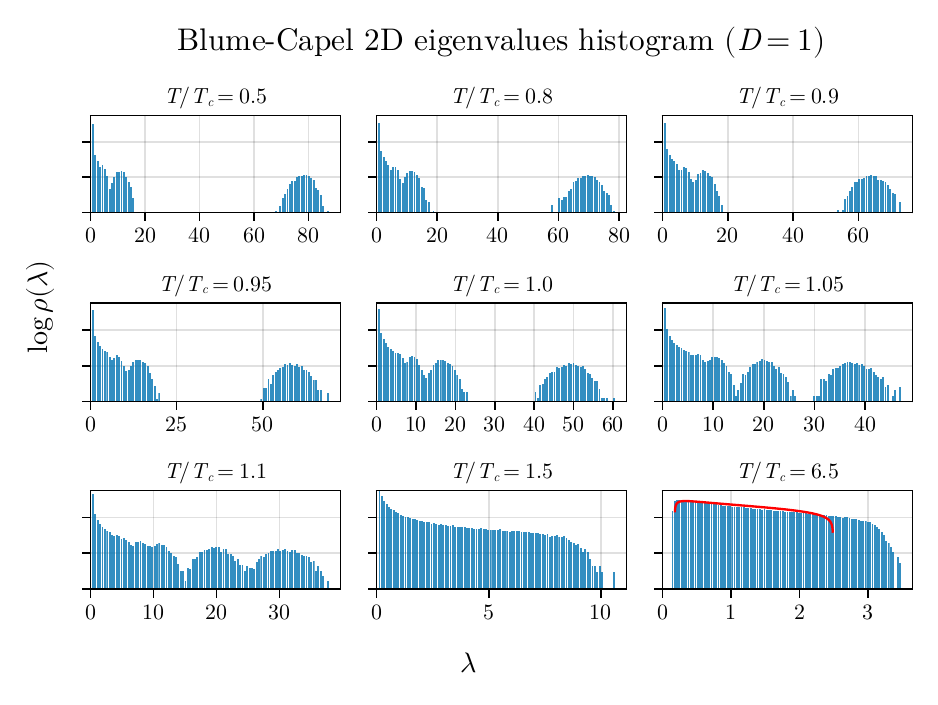}
\end{center}
\caption{The Density of States in the Two-Dimensional BC Model with
anisotropy ($D=1$). The gap between eigenvalues varies with the temperature
of the simulated system. While the system approaches the MP law, an exact
match is not achieved at high temperatures ($T>T_{C}$) due to the presence
of spin-spin correlations, preventing complete correspondence.}
\label{Fig:2D_BC_DE}
\end{figure*}

We begin our results by displaying the histogram of eigenvalues. We choose $%
D=1$ for both the 2D and 3D BC models to illustrate the density of
eigenvalues obtained through the diagonalization of matrices $G^{\ast}$.
Figure \ref{Fig:2D_BC_DE} presents histograms for various temperatures. An
evolution of the gap between the two eigenvalue bulks can be observed.
Similar behavior is noted for the Ising model on two-dimensional lattices
under mean-field approximations \cite{RMT2023, RMT-2}.

\begin{figure*}[tbp]
\begin{center}
\includegraphics[width=1.0%
\columnwidth]{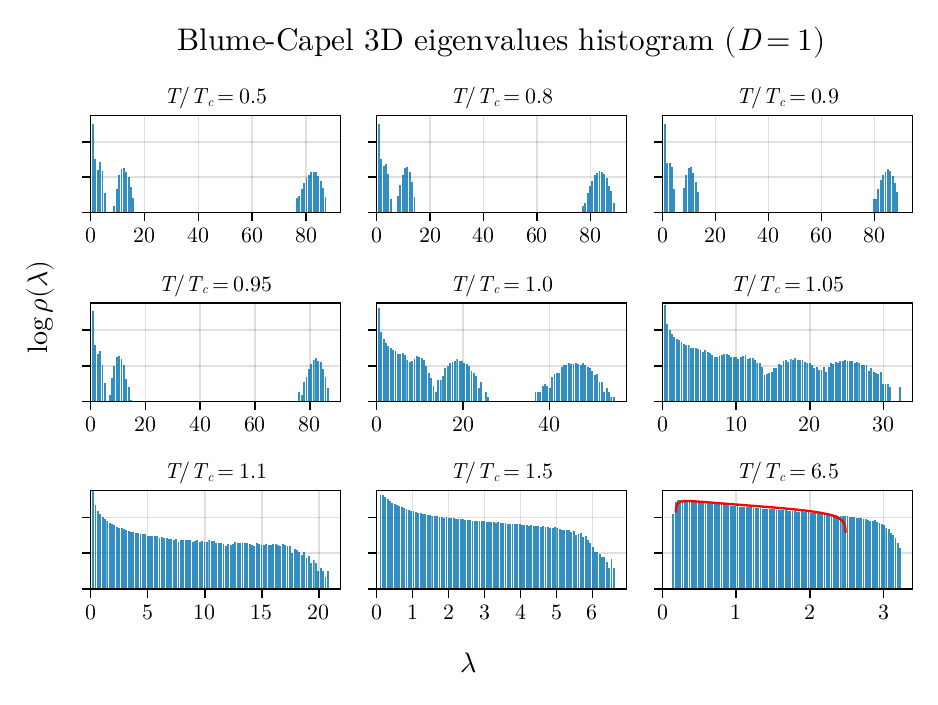}
\end{center}
\caption{The density of states for anisotropy $D=1$ in the three-dimensional
BC model. Similar behavior in the gap between two bulk eigenvalues is
observed compared to the two-dimensional BC model (see Figure \protect\ref%
{Fig:2D_BC_DE}) }
\label{Fig:3D_BC_DE}
\end{figure*}

We can discern analogous behavior in the three-dimensional BC model, as
illustrated in Figure \ref{Fig:3D_BC_DE}.

It's crucial to highlight the distinctive trend of the eigenvalue gap
narrowing around the critical temperature, along with the correspondence to
the MP law for $T > T_{C}$. While a perfect agreement is not
expected as $T$ approaches infinity due to the correlation matrix's
construction, incorporating total magnetization and time series with
inherent autocorrelation, it's important to note that this doesn't diminish
the method's validity in any manner.

\begin{figure*}[tbp]
\begin{center}
\includegraphics[width=1.0%
\columnwidth]{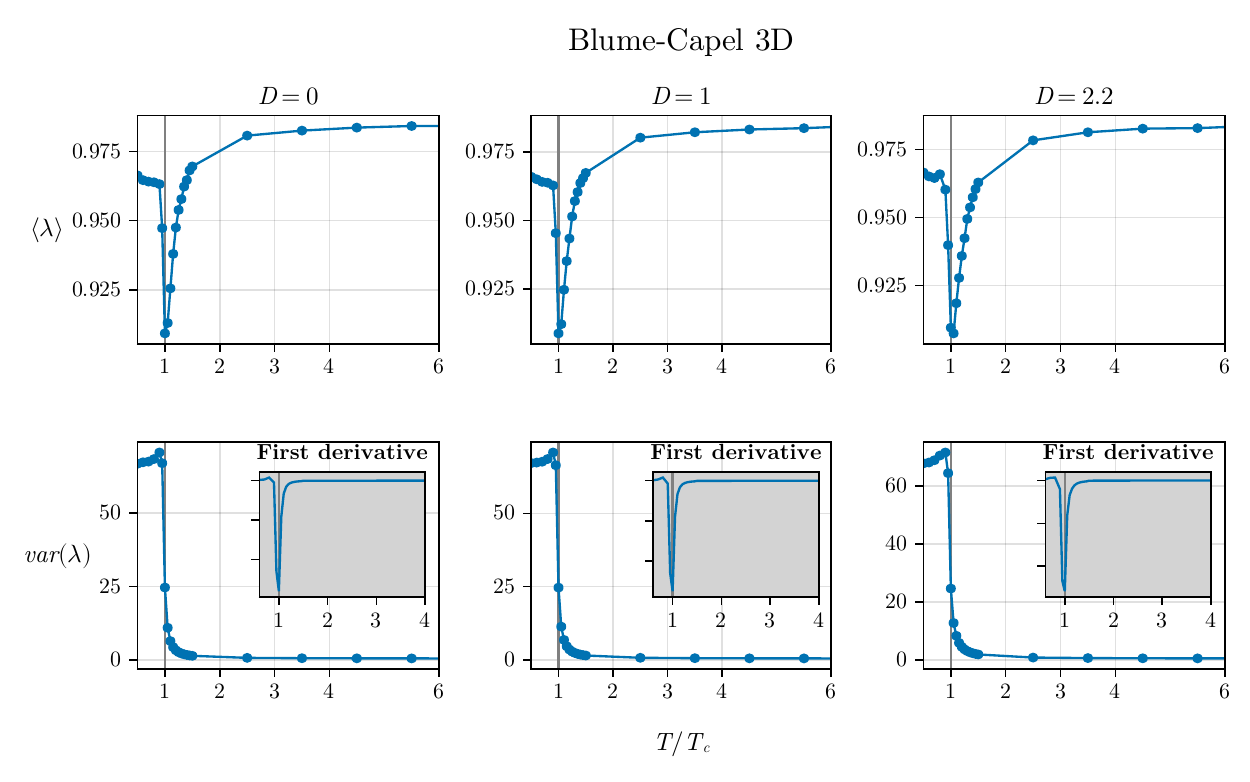}
\end{center}
\caption{The average and variance of the three-dimensional BC model as a
function of temperature illustrate a similar behavior occurring in three
dimensions. The inset plots depict the derivative of the variance,
highlighting its divergence at $T=T_{C}$. Interestingly, it is observed that
the inflection point appears to be even more pronounced in three dimensions.}
\label{Fig:fluctuations_3D}
\end{figure*}

However, it is necessary to utilize this density of states to effectively
determine the critical parameter. This can be achieved by computing the
moments of the density of states, specifically $\left\langle \lambda
\right\rangle $ and $\left\langle \left( \Delta \lambda \right)
^{2}\right\rangle $. In this regard, we observe the results for three
different values of $D$. For the two-dimensional BC model (refer to Figure \ref%
{Fig:fluctuations_2D}), we tested three values: $D=0$, $D=1$, and $D=1.75$,
employing the corresponding $T_{C}$ values estimated in \cite{Butera} as a
basis. Similarly, for the 3D BC model, we utilized $D=0$, $D=1$, and $D=2.2$,
as depicted in Figure \ref{Fig:fluctuations_3D}.

\begin{figure*}[tbp]
\begin{center}
\includegraphics[width=1.0%
\columnwidth]{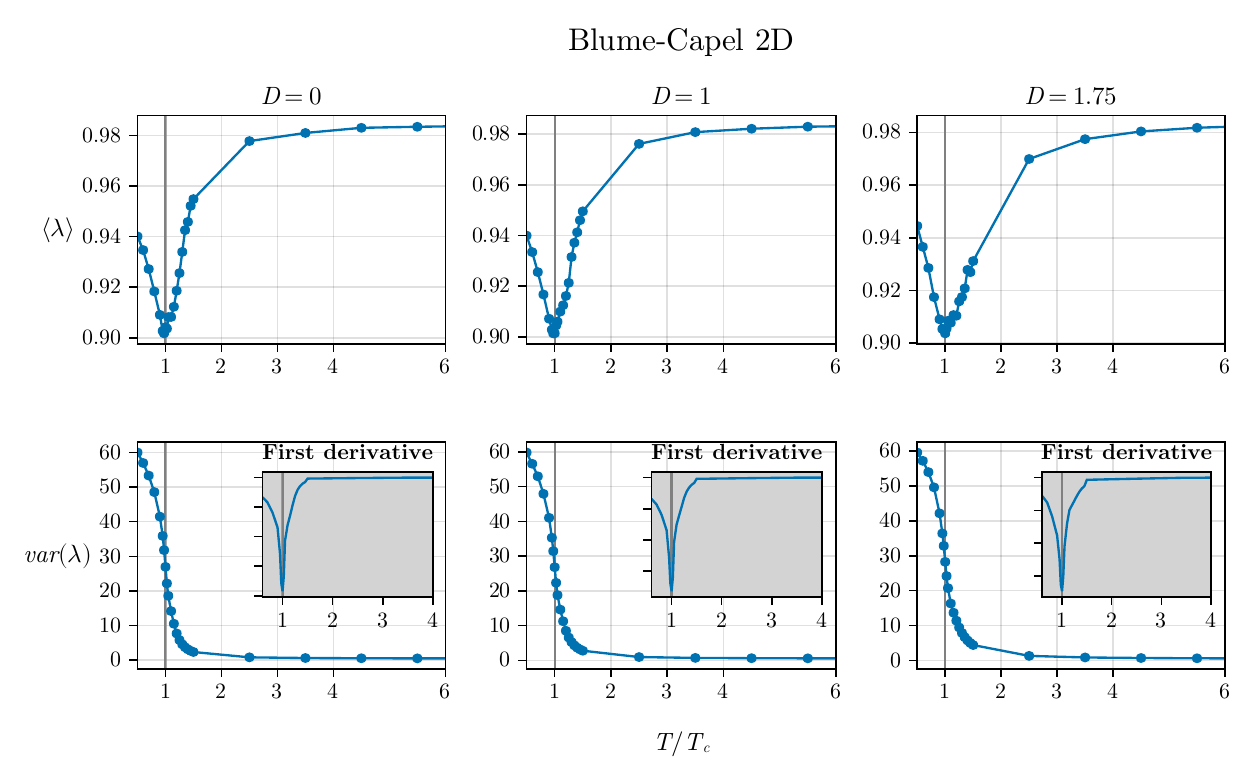}
\end{center}
\caption{Average and variance of the two-dimensional BC model as a function
of temperature are depicted. The inset plots show the derivative of the
variance, indicating a divergence at $T=T_{C}$.}
\label{Fig:fluctuations_2D}
\end{figure*}

We can observe a pronounced minimum in $\left\langle \lambda \right\rangle$
at $T=T_{C}$ in both the two-dimensional and three-dimensional versions of
the BC model, which is related to the closing gap observed in Figures \ref%
{Fig:2D_BC_DE} and \ref{Fig:3D_BC_DE}. Additionally, an inflection point
seems to be observed for the variance exactly at $T=T_{C}$ in both versions
of the model (in two and three dimensions), demonstrating that both spectral
measures -- the average and variance -- are effective in exploring
criticality. The inset plot displays the first derivative of the variance:

\begin{equation*}
\alpha =\frac{d\left\langle \left( \Delta \lambda \right) ^{2}\right\rangle 
}{dt}=T_{C}\frac{d\left\langle \left( \Delta \lambda \right)
^{2}\right\rangle }{dT}\text{,}
\end{equation*}%
indicating that the critical temperature is associated with a pronounced
minimum (a negative value of significant magnitude), where $t = \frac{T}{T_{C}}$

\begin{figure*}[tbp]
\begin{center}
\includegraphics[width=1.0%
\columnwidth]{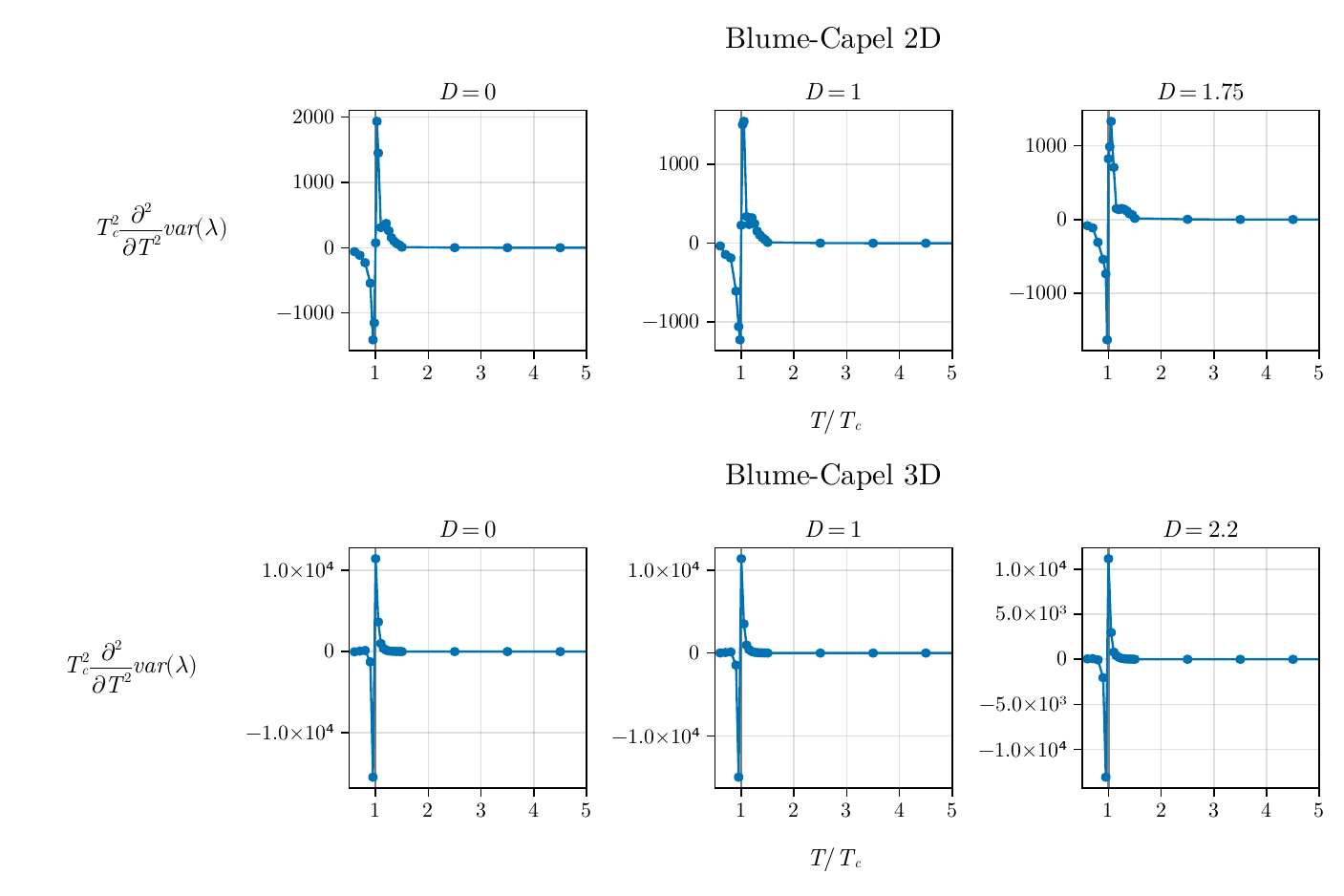}
\end{center}
\caption{Second derivative of variance ($\protect\zeta$) for both the
two-dimensional and three-dimensional BC models. The critical temperature
precisely corresponds to the inflection point of the eigenvalue variance,
indicated by $\protect\zeta <0$ for $T<T_{C}$ and $\protect\zeta >0$ for $%
T>T_{C}$.}
\label{Fig:corroboration_inflection_point}
\end{figure*}

To better understand such behavior, we examine the second derivative: 
\begin{equation*}
\zeta =\frac{d^{2}\left\langle \left( \Delta \lambda \right)
^{2}\right\rangle }{dt^{2}}=T_{C}^{2}\frac{d^{2}\left\langle \left( \Delta
\lambda \right) ^{2}\right\rangle }{dT^{2}}
\end{equation*}%
and its plot is depicted in Figure \ref{Fig:corroboration_inflection_point}
for both scenarios: the two-dimensional and three-dimensional BC models.

We notice that the critical temperature precisely aligns with the inflection
point of the eigenvalue variance due to the condition $\zeta <0$ for $%
T<T_{C} $ and $\zeta >0$ for $T>T_{C}$. Understanding the nature of this
inflection point warrants further investigation, prompting a thorough
discussion. In our work, we provide an in-depth analysis of this aspect in
the appendix.

Thus, in this first subsection, we observed that critical points of the BC
model are well captured by this spectral methodology in both versions of the
model: two and three dimensions. We used different parameters based on
fluctuations of the eigenvalues and their convenient derivatives to conduct
our analysis. Now it is important to utilize this method to explore some
nuances of points near the tricritical one. We will demonstrate how the
method responds to the crossover effect.

\subsection{Crossover phenomena}

We begin by simulating the average eigenvalue as a function of $T/T_{C}$.
However, our focus now shifts to examining points near the tricritical point
(TCP) to observe how the spectra of Wishart matrices behave when approaching
this point alongside time series of magnetization simulated with (MC) simulations.
By repeating our procedure, we initially investigate the
issue in two dimensions to understand how the spectrum respond to the
expected crossover phenomena in this model (refer to Figure \ref%
{Fig:Crossover_2D_average}). To accomplish this, we employed the values $%
D=1.9$, $1.92$, $1.9336$, $1.9421$, $1.9501$, and $1.96582$ (TCP).

\begin{figure*}[tbp]
\begin{center}
\includegraphics[width=1.0%
\columnwidth]{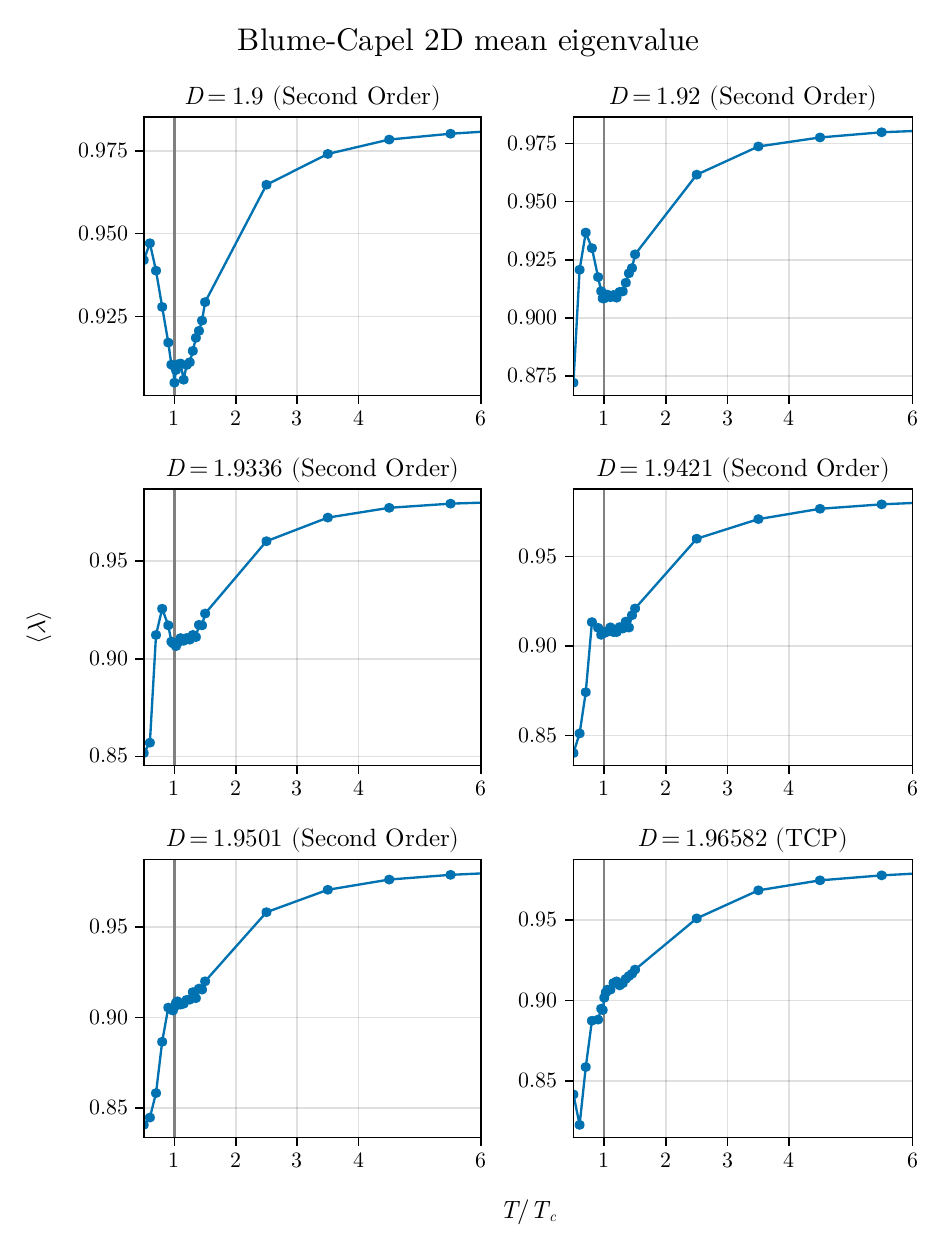}
\end{center}
\caption{Average eigenvalue approaching the TCP in the
two-dimensional BC model. We can observe that the shape of the curve is
deformed as we approach the TCP on the critical line.}
\label{Fig:Crossover_2D_average}
\end{figure*}

We notice that the minimum becomes less pronounced and deformed as we
approach the TCP. However, it is interesting to note that even for points
near the TCP, the method indicates the critical point albeit with reduced
precision. Initially, the peak transforms into a shell, resembling a
shoulder, and eventually, at the tricritical point, the minimum completely
disappears.

\begin{figure*}[tbp]
\begin{center}
\includegraphics[width=0.9%
\columnwidth]{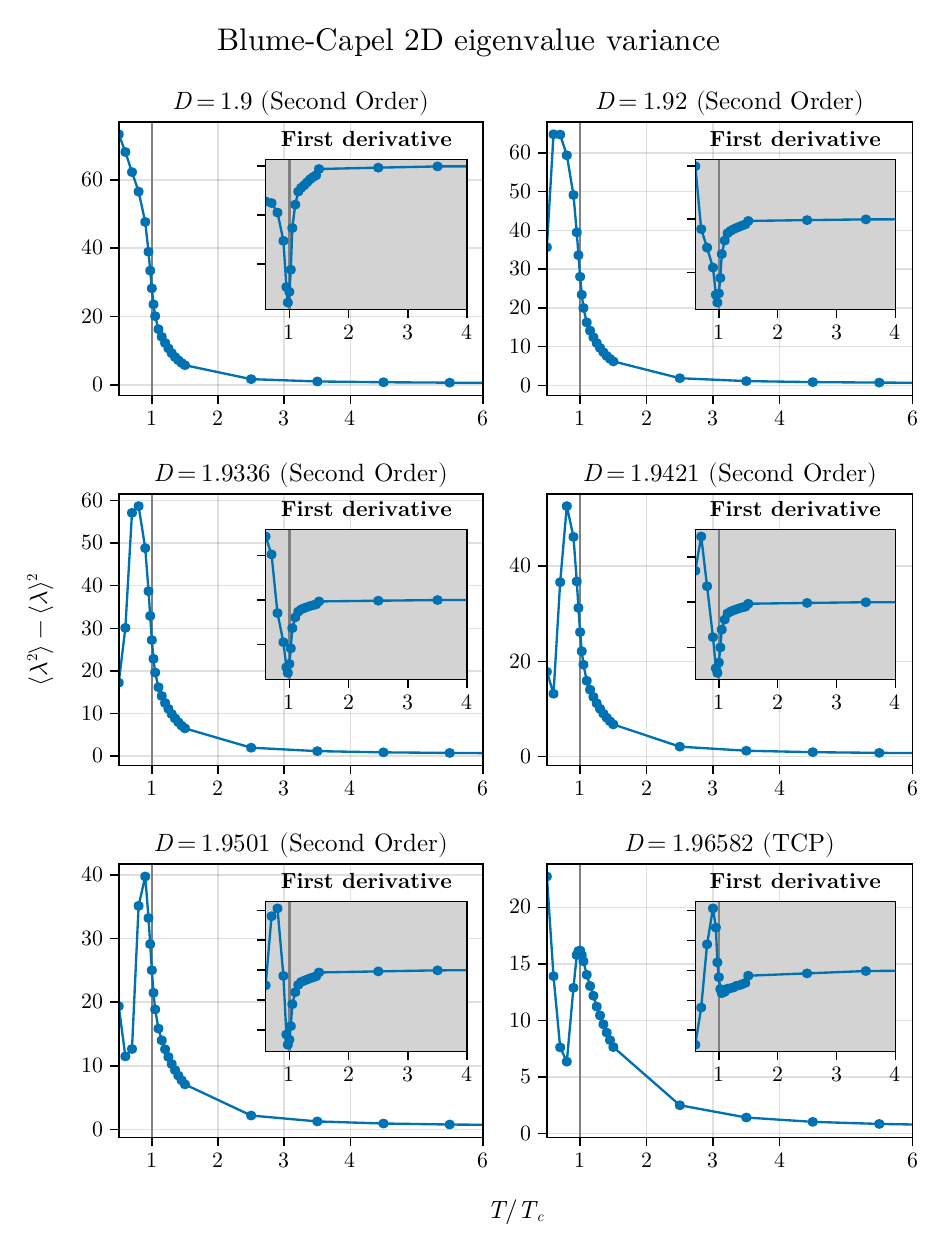}
\end{center}
\caption{Eigenvalue variance as a function of temperature approaches the TCP
in the 2D-BC model. The method appears to reasonably
respond even for points closer to the TCP. We can observe the inflection
point up to just before the TCP, but we also notice a small deviation between
the critical exact values and those determined by the method due to the
crossover. At this precise TCP, there is a peak at the tricritical
temperature that shifts from the previous points. Interestingly, at the TCP,
we do not observe the inflection point in two dimensions.}
\label{Fig:Crossover_2D_variance}
\end{figure*}

This indicates that the average, which localized the critical points well
away from the TCP, strongly suffers the influence of this point, showing that
the spectra of our correlation matrices precisely reflects what occurs with
the thermodynamics of the model. Following this, we observe the dispersion
of eigenvalues. We plot $\left\langle \left( \Delta \lambda \right)
^{2}\right\rangle $ as a function of $T/T_{C}$ for the same values of $D$
previously used to study $\left\langle \lambda \right\rangle $. This result
is presented in Figure \ref{Fig:Crossover_2D_variance}.

In contrast to the behavior observed with $\left\langle \lambda
\right\rangle $, the quantity $\left\langle \left( \Delta \lambda \right)
^{2}\right\rangle $ exhibits an inflection point very close to $T=T_{C}$,
even for points near the TCP; i.e., the variance senses the crossover but is
not completely extinguished as with the simple average.

However, at this precise juncture, a peak occurs at the TCP, which, upon
closer examination, appears to shift as $D$ approaches $D_{TCP}$, i.e., we
observe a migration of the maximum that will coincide at the critical
temperature only exactly at TCP. Particularly intriguing is the observation
that the migration of the maximum occurs with a decrease in its amplitude as 
$D$ approaches $D_{TCP}$.

\begin{figure*}[tbp]
\begin{center}
\includegraphics[width=1.0%
\columnwidth]{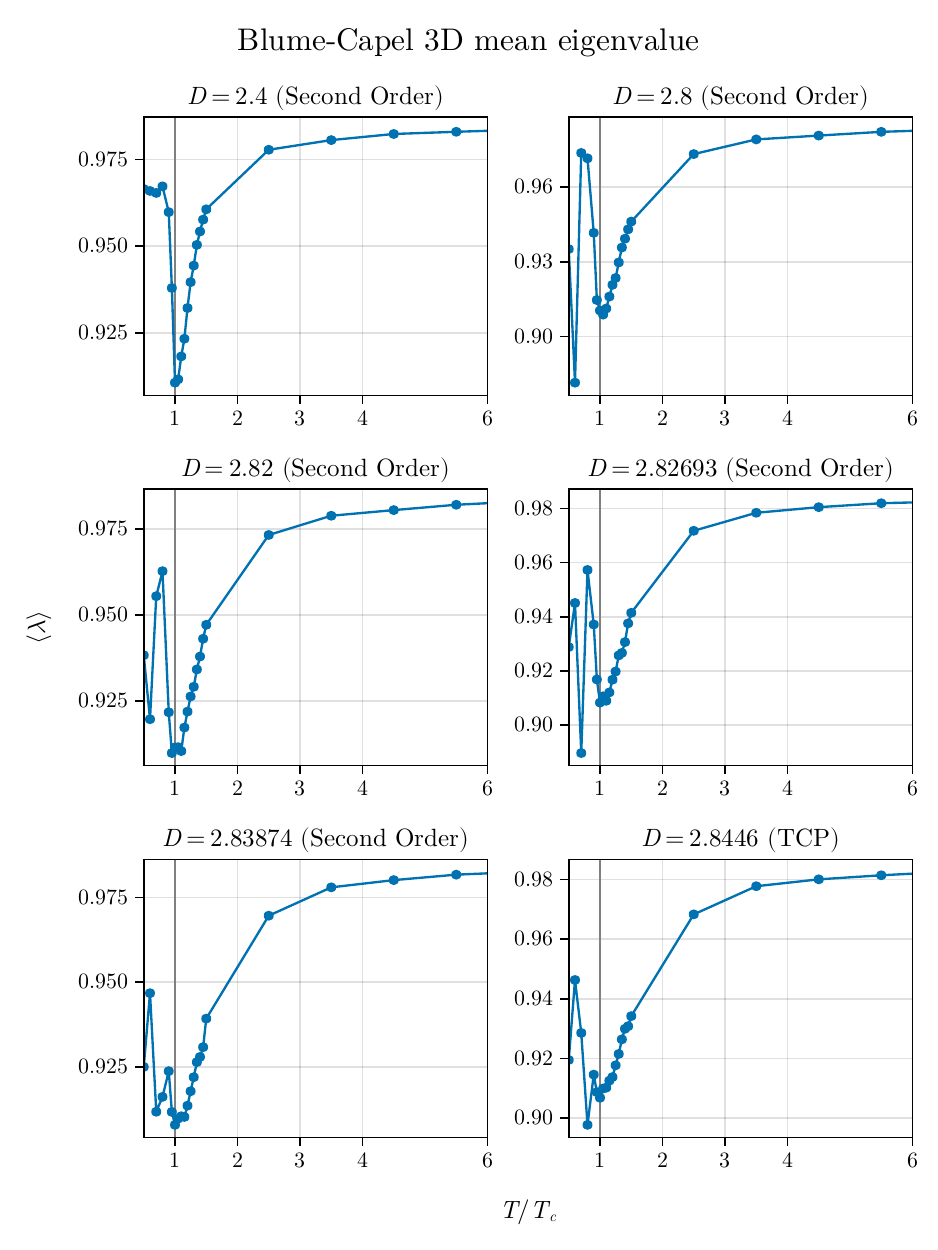}
\end{center}
\caption{Average eigenvalue approaching the TCP in the 3D BC model, mirroring
the analysis conducted for the 2D version shown in Figure \protect\ref%
{Fig:Crossover_2D_average}. }
\label{Fig:Crossover_3D_average}
\end{figure*}

Now, we extend this investigation to the three-dimensional BC model. The
behavior of the average as a function of $T/T_{C}$ for different values of
$D$ is illustrated in Figure \ref{Fig:Crossover_3D_average}.

Exactly as occurred in the two-dimensional version of the model, the method
indicates the critical point but loses precision as it approaches the TCP.
However, we observe that for the 3D version, the minimum is more persistent,
even at the TCP itself, since we do not observe a shoulder as obtained in the
two-dimensional version.

Here, it is important to mention that in the three-dimensional version of
the BC model, the short-time regime presents a logarithmic correction \cite%
{Janssen1994,SilvaBC3D2022}, which should suggest such different behavior.
However, again, for points far from the TCP, the minimum of $\left\langle
\lambda \right\rangle $ always occurs at $T=T_{C}$ as the method exactly
prescribes in its original proposal.

\begin{figure*}[tbp]
\begin{center}
\includegraphics[width=1.0%
\columnwidth]{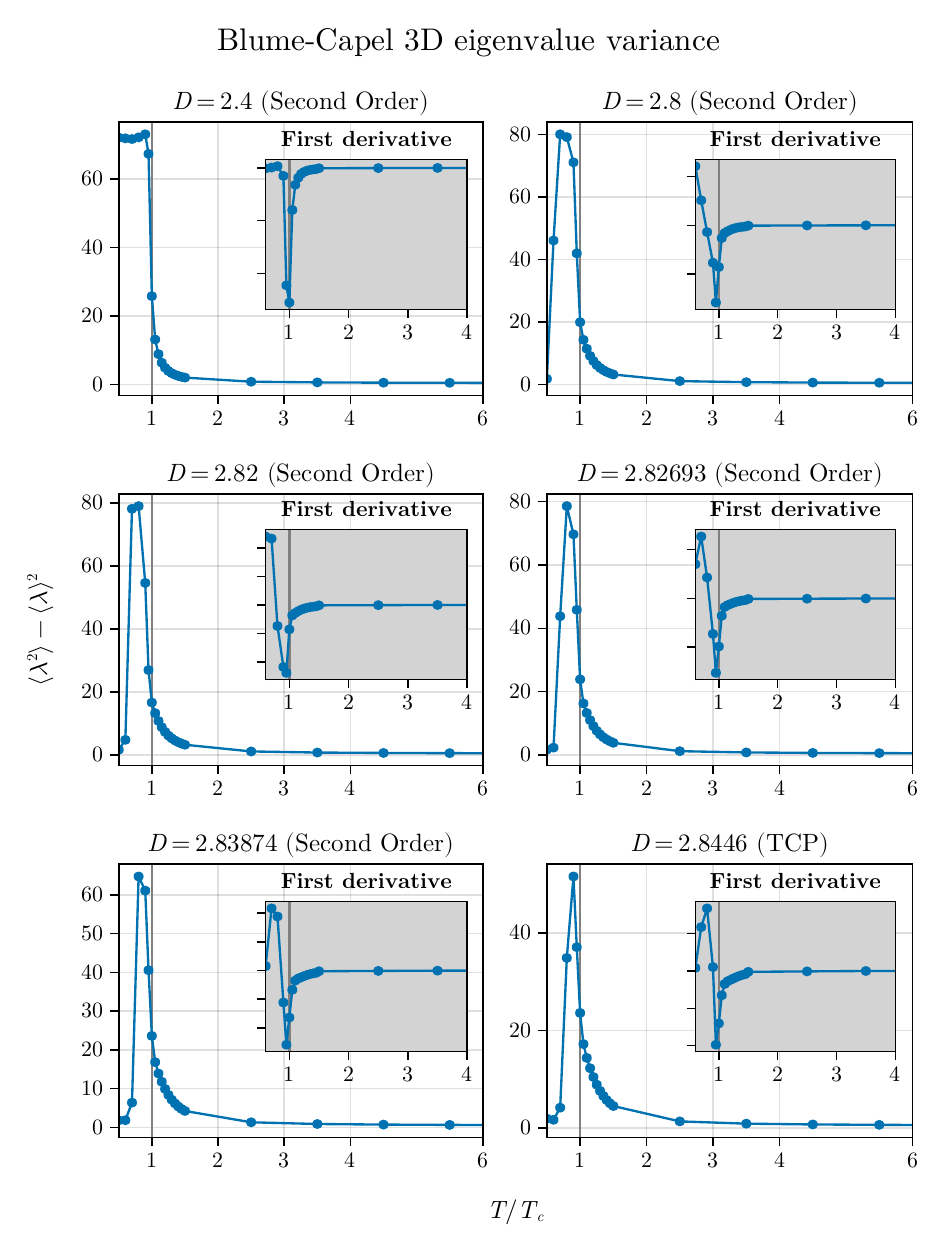}
\end{center}
\caption{Eigenvalue variance approaching the TCP in the 3D-BC model. We can
observe the inflection point until the TCP point, but that slightly differs
from the best estimates of the critical temperatures in this vicinity of the
TCP.}
\label{Fig:Crossover_3D_variance}
\end{figure*}

And what about the variance? Similar to what occurred in the two-dimensional
version, the inflection point appears for all points. Their estimates in the
vicinity of the TCP point slightly differ from the exact critical value. This
further reinforces and suggests that we can use the inflection point of the
spectral variance as a reliable indicator of critical phenomena (see Figure %
\ref{Fig:Crossover_3D_variance}) far from TCP, but crossover effects can
generate small deviations around TCP.

Crossover effects are observed in several works, and they play an important
role in determining other quantities related to critical behavior, such as
critical exponents. Here, we study their influence on the spectra of Wishart
matrices built with time series of magnetization of the BC model.

Our results suggest that $\left\langle \lambda \right\rangle $ works very
well for critical points outside the influence of the crossover, but it
is not a good indicator of criticality near the TCP. In this case, we can
make use of eigenvalue variance, which exhibits an inflection point at the
critical temperature and responds reasonably well even when near the TCP,
although it is also sensitive to crossover effects.

It is important to mention that MC simulations, whether
in equilibrium or nonequilibrium, are generally sensitive to crossovers. For
example, the dynamic exponent $z$, expected to be universal, is
significantly influenced along the critical line in two dimensions \cite%
{Silva2002,Silva2013}, and even in the mean-field regime \cite{SilvaBJPMF}.

In Statistical Mechanics, the role of the maximum eigenvalue appears in many contexts,
and an important question is whether they can also be used to quantify critical phenomena in the spectral
method developed here. In other words, does the maximum eigenvalue of
Wishart matrices respond to the critical behavior of the BC model? The
answer is positive, and we will present the results in the next subsection.

\subsection{Analyzing Extreme Statistics of Correlation Magnetization
Matrices}

The utilization of extreme values has been extensively investigated within formal 
contexts to characterize phase transitions in random matrices (see, for instance, \cite{Baik2005,Benaych2011}). Nevertheless, we posit that our approach holds promise 
for extension, leveraging similar principles computationally and efficiently to pinpoint 
critical points. Hence, this paper embarks on an exploration of extreme value statistics 
as indicators of critical points within the BC model, employing our correlation magnetization 
matrices.

\begin{figure*}[tbp]
\begin{center}
\includegraphics[width=1.0\columnwidth]{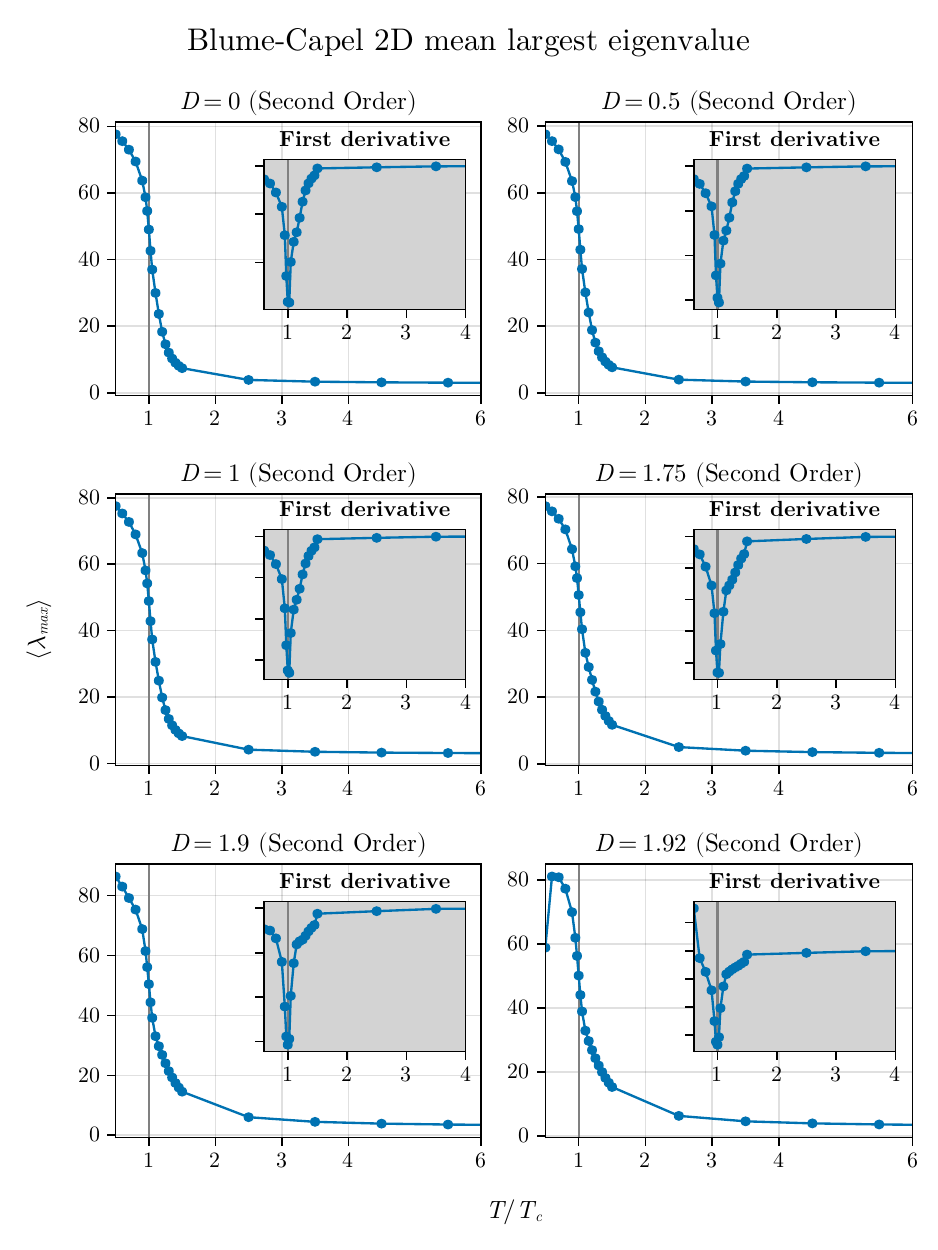}
\end{center}
\caption{Averaged maximum eigenvalue as a function of $T/T_{C}$ for different values 
of $D$ in the two-dimensional BC model. }
\label{Fig:Maximum_eigenvalue_2D}
\end{figure*}

\begin{figure*}[tbp]
\begin{center}
\includegraphics[width=1.0\columnwidth]{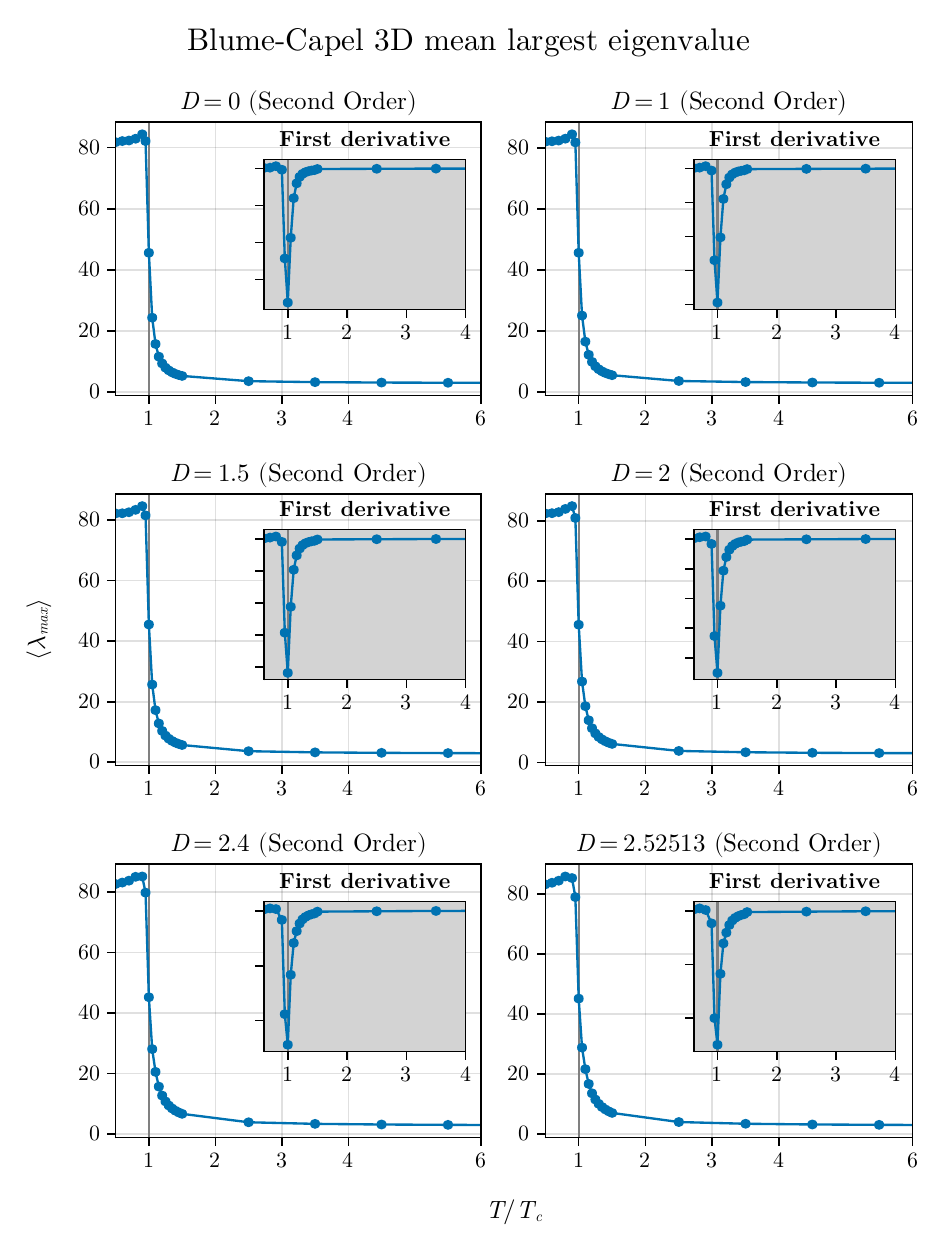}
\end{center}
\caption{Averaged maximum eigenvalue as function of $T/T_{C}$ for different
values of $D$ in the three-dimensional BC model. }
\label{Fig:Maximum_eigenvalue_3D}
\end{figure*}

Accordingly, for each matrix $G^{\ast }$ constructed, we extract its maximum eigenvalue and 
compute the average across multiple runs using the following formula:%
\begin{equation*}
\left\langle \lambda _{\max }\right\rangle =\frac{1}{N_{run}}%
\sum\limits_{i=1}^{N_{run}}\lambda _{\max }^{(i)}
\end{equation*}%
and consider its behavior as a function of different temperatures for the BC model in 
both two and three dimensions. For such analysis, we choose $D=0$, $0.5$, $1.0$, $1.75$, $1.9$, 
and $1.92$ in two dimensions and $D=0$, $1$, $1.5$, $2$, $2.4$, and $2.52513$ for the 
three-dimensional version of the model. The behavior of $\left\langle \lambda_{\max }\right\rangle$%
as a function of $T/T{C}$ is shown in Figures \ref{Fig:Maximum_eigenvalue_2D} and \ref{Fig:Maximum_eigenvalue_3D} respectively for the cases of the two and three-dimensional 
BC models.

We observe that in both situations, the critical point is identified by a notable inflection point. Additionally, we show the first derivative in relation to $\frac{T}{T_{C}}$, simply described as $T_{C}\frac{d\left\langle \lambda_{\max }\right\rangle }{dT}$, as a function of $\frac{T}{T_{C}}$ as inset plots in these figures.

Thus, we can also observe that the averaged maximum eigenvalue responds to the 
criticality of the system for the different critical points studied here for the BC model 
regardless of dimensionality. This adds an additional parameter to our framework to 
identify criticality in spin systems that can be tested in other models. In the next section, 
we will conclude our analysis by showing that the method that uses short times (in this current 
contribution and in the previous ones \cite{RMT-2,RMT2023} we used $N_{MC}=300$ steps) 
also works with small systems. Up to now, we have used $L=100$ in two dimensions. 
We will demonstrate that this number can be further reduced.

\subsection{Finite Size Scaling: Exploring Small Systems with Short Time Scales}

The method's efficiency in saving computer time through the use of short time scales presents a particularly intriguing prospect. For instance, in this current study, we employed $N_{MC}=300$ steps. Therefore, to highlight the versatility of our method, we will explore another aspect: the system size. We have investigated this aspect in both two and three dimensions, demonstrating that systems can be studied effectively with even smaller sizes, yielding good estimates.

\begin{figure*}[tbp]
\begin{center}
\includegraphics[width=1.0\columnwidth]{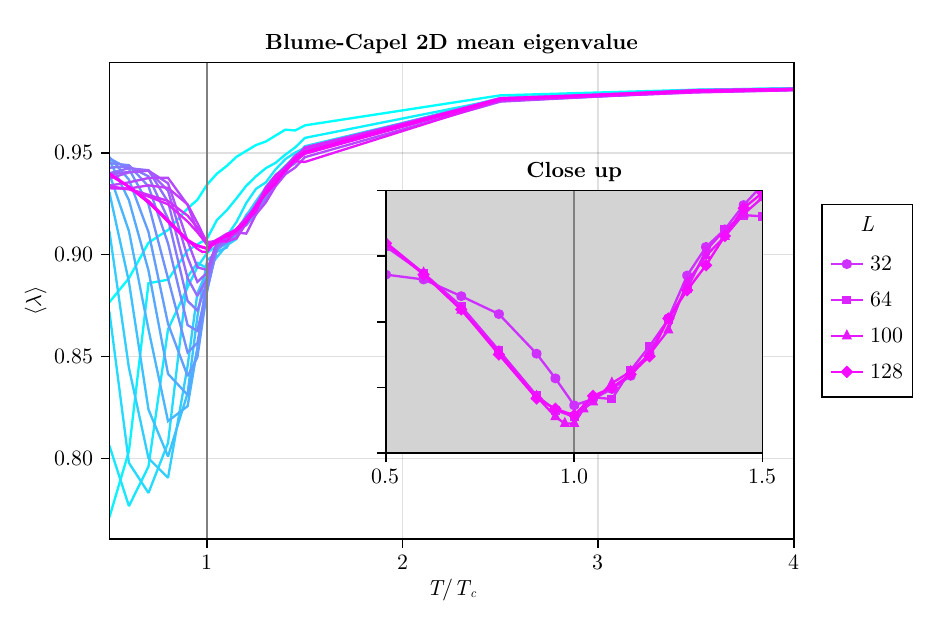}
\end{center}
\caption{The average eigenvalue as a function of $T/T_{C}$ for different linear 
system sizes is depicted. We start with $L=2,3,4,...,16$, then proceed to larger sizes 
including $L=20$, $25$, $30$, $32$, $64$, $100$, and $128$ for the two-dimensional BC 
model with $D=0$, chosen for simplicity. The inset plot illustrates that for $L\geq 32$, the 
minimum at $T=T_{C}$ coincides. With $L\geq 64$, there is excellent agreement.}
\label{Fig:FSS_2D}
\end{figure*}

We deliberately selected only the case of $D=0$ without loss of generality. The average eigenvalue is plotted as a function of $T/T_{C}$ for different sizes of the two-dimensional BC system (refer to Figure \ref{Fig:FSS_2D}). Initially, we explore sizes ranging from $L=2$ to $L=16$, and subsequently extend to $L=20$, $25$, $30$, $32$, $64$, $100$, and $128$.

We can observe an influence of the system size, where for small systems, a 
minimum at the exact $T_{C}$ is found. The inset plot illustrates that for $L\geq 32$, 
the minimum at $T=T_{C}$ coincides. With $L\geq 64$, there is excellent agreement.

For the three dimensional model, our investigation yields similar results.
Encouragingly, we found consistent behavior, particularly noteworthy for $L \geq 16$,
where a distinct trend emerges: the average eigenvalue reaches a minimum precisely at $T=T_{C}$.
This observation implies the feasibility of exploring intricate phenomena within compact systems.
Thus, the potential for fruitful spectral analyses in modest-scale systems becomes increasingly evident.

\begin{figure*}[tbp]
\begin{center}
\includegraphics[width=1.0\columnwidth]{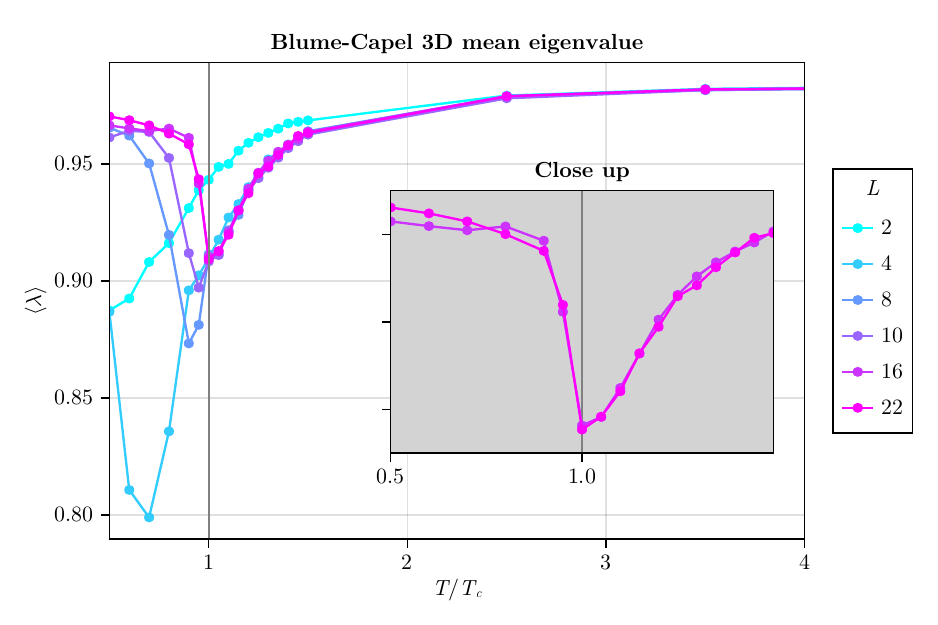}
\end{center}
\caption{Average eigenvalue plotted against $T/T_{C}$ for various linear system 
sizes. We consider $L=2$, $4$, $8$, $10$, $16$, and $22$ in a three-dimensional BC 
model with $D=0$ for simplicity. The inset plot highlights that for $L\geq 16$, the 
minimum occurs precisely at $T=T_{C}$. }
\label{Fig:FSS_3D}
\end{figure*}

As we wrap up this subsection, it's remarkable to note that in addition to employing 
short-time evolution of magnetization ($N_{MC}=300$ steps), we can also leverage small systems to achieve robust results. Surprisingly, for $L\geq 64$ in the 
two-dimensional BC model and $L\geq 16$ in the three-dimensional BC model, our 
method accurately identifies the critical temperature of the model. This underscores the efficacy of the spectral method, highlighting its strength in pinpointing critical points.

Readers are encouraged to juxtapose our method with simpler MC simulations, considering its additional workload in terms of matrix diagonalization. However, it operates on low-dimensional matrices (with $N_{sample}=100$ here -- adjustable for further optimization). In comparison to time-dependent simulations, which may require, for instance, $L=256$ and a substantial number of runs to adequately sample quantities (with a minimum of $2000$ runs for ferromagnetic initial states and over $10^{4}$ for disordered initial states where $m_{0}\approx 0$), our method presents an intriguing alternative. It's worth noting that capturing the thermodynamics of the model with such accuracy using a computationally "cheap" spectral method is not a trivial achievement.  

Equilibrium MC simulations are plagued by the issue of critical slowing down, compounded by the use of larger lattices than those employed in our approach. While a thorough comparison between this spectral method and standard MC methods warrants attention, we must emphasize the compelling observation that the thermodynamics of the systems are remarkably well-reflected by this "spectral thermodynamics".

An avenue ripe for exploration is the investigation of long-range systems, which will undoubtedly command our focus in future applications, precisely due to the lack of requirement for large-scale systems.

\section{Conclusions}

In this study, we have extended a method originally developed in \cite{RMT2023} to describe the spin-1 Ising model with anisotropy, known as the Blume-Capel model. This model exhibits a tricritical point in both two and three dimensions. Our method has proven effective in accurately capturing these critical points and illustrating the associated crossover phenomena.

Furthermore, we underscore the computational efficiency of our proposed method compared to similar approaches. By diagonalizing matrices of size $O(N_{sample})$, where $N_{sample}$ is set to 100 in this work, we alleviate the computational burden. This stands in contrast to other spectral methods in the literature, which necessitate diagonalizing matrices of size $O(L^{d})$, where $L$ represents the linear dimension of the system and $d$ its dimensionality. For instance, in a system with $L=100$ and $d=2$, this would entail diagonalizing matrices of size $10^{4}\times 10^{4}$ for a significant number of runs, which is computationally intensive.

In summary, our findings demonstrate that spectral methods provide a promising avenue for characterizing the thermodynamics of spin systems exhibiting tricritical points and crossover phenomena, regardless of the system's dimensionality. Notably, we achieved these results using very small systems and short time series, suggesting a means to bypass both critical slowing down and the necessity for extremely large systems often observed in standard MC simulations that do not involve the diagonalization of Wishart matrices.

While our proposal does not seek to directly compete with standard MC simulations, either in equilibrium or nonequilibrium settings, our results indicate that the method merits consideration for application in these contexts due to its efficiency and sensitivity.

Lastly, we emphasize the remarkable success of the developed model in characterizing chaos \cite{SilvaDynamics}, as well as in describing the aging effects in spin systems \cite{PRLnew}, underscoring the ongoing exploration of its full potential and the depth of understanding yet to be achieved in this research domain. 

\section{Appendix}

\begin{figure*}[tbp]
\begin{center}
\includegraphics[width=1.0\columnwidth]{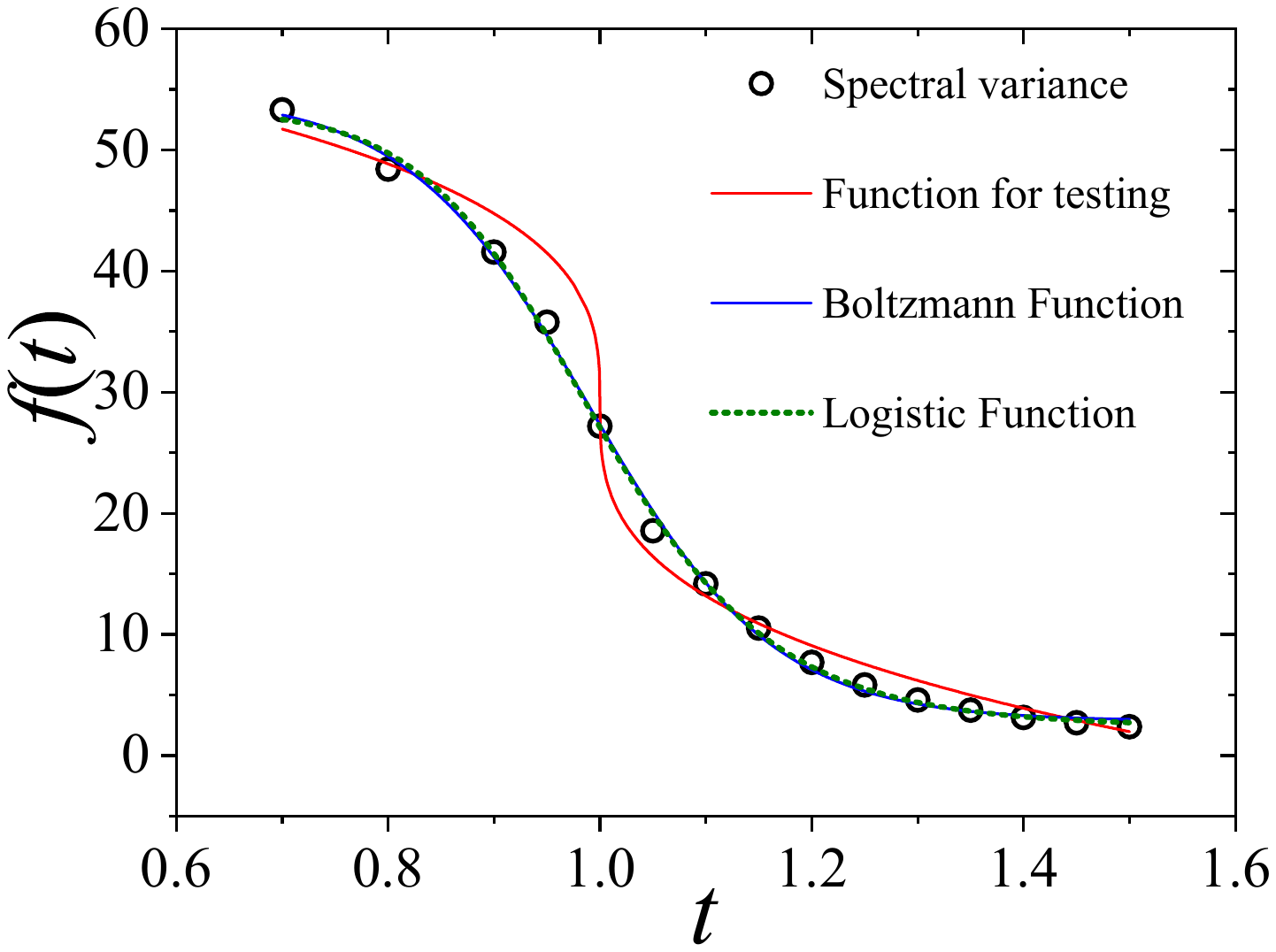}
\end{center}
\caption{Spectral variance as a function of various values of $t=T/T_{C}$ for $D=0$, including $t=1$, is presented. We provide fits using the Boltzmann, logistic, and a function exhibiting a unique type of inflection point at $t=1$. }
\label{Fig:fits}
\end{figure*}

We observed that the inflection point appearing in the spectral variance serves as an 
excellent indicator of critical behavior when employing the method of constructing Wishart matrices.
Clearly, within this framework, we can explore the nature of the inflection point under consideration.
As a preliminary test, we opted for the simple case $D=0$ and plotted the spectral variance for various values of $t=T/T_{C}$, including $t=1$, as depicted in Figure \ref{Fig:fits}. We fitted two suggestive functions to the data, beginning with the well-known logistic function:%
\begin{equation*}
var(t,t_{0})=\frac{c_{1}}{1+\left( \frac{t}{t_{0}}\right) ^{p}}+c_{2}
\end{equation*}%
and the second one, the Boltzmann function:%
\begin{equation*}
var(t,t_{0})=\frac{c_{1}}{1+\exp \left[ \left( \frac{t-t_{0}}{p}\right) %
\right] }+c_{2}\text{.}
\end{equation*}%
Both fits were conducted by initially fixing $t_{0}=1$, as $c_{1}$ and $c_{2}$ represent normalization and fitting parameters, respectively. Thus, the fit essentially revolves around one parameter: $p$. This observation arises from the fact that both functions can be expressed as:%
\begin{equation*}
var(t,t_{0})=c_{1}\varphi (t)+c_{2}
\end{equation*}%
where $\varphi (t)=\frac{1}{1+t^{p}}$ for the logistic function and $\frac{1}{1+\exp %
\left[ \left( \frac{t-1}{p}\right) \right] }$ for the Boltzmann function. We can
observe a good visual fit, both with a coefficient of determination $r\approx
0.998$. For additional information regarding the fits presented in Figure \ref{Fig:fits}, for the logistic function we obtained: $c_{1}=50.4\pm 0.8$, $c_{2}=2.3\pm 0.4$, and $%
p=11.8\pm 0.5$. Similarly, for the Boltzmann function, we obtained: $c_{1}=52\pm 1$, $c_{2}=2.8\pm 0.4$, and $%
p=0.087\pm 0.004$.

First, we would like to demonstrate that $t_{0}=1$ serves as an inflection point for both functions. In the first case, we have:
\begin{equation*}
\frac{d^{2}\varphi }{dt^{2}}=\frac{p(p-1)}{(1+t^{p})^{3}}t^{p-2}(t^{p}-1)
\end{equation*}
and it is certain that $p>1$ for this type of curve. Thus, for $t>1$, $\frac{d^{2}\varphi }{dt^{2}}>0$, while for $t<1$, $\frac{d^{2}\varphi }{dt^{2}}<0$, demonstrating that the Boltzmann function with $p>1$ has an inflection point at $t=1$. Specifically, $\left. \frac{d^{2}\varphi }{dt^{2}}\right\vert _{t=1}=0$.

Similarly, for the logistic function, we have: 
\begin{equation*}
\frac{d^{2}\varphi }{dt^{2}}=\frac{1}{p^{2}}\frac{e^{\frac{t-1}{p}}}{(e^{%
\frac{t-1}{p}}+1)^{3}}(e^{\frac{t-1}{p}}-1)
\end{equation*}

For $t>1$, $e^{\frac{t-1}{p}}>1$, and therefore $\frac{d^{2}\varphi }{dt^{2}}%
>0$, and for $t<1$, $e^{\frac{t-1}{p}}<1$, and $\frac{d^{2}\varphi }{dt^{2}}%
<0$. This shows that $t=1$ is also an inflection point. Again $\left. \frac{%
d^{2}\varphi }{dt^{2}}\right\vert _{t=1}=0$.

In the same Figure \ref{Fig:fits} we purposively presented an additional fit
to a function: 
\begin{equation*}
var(t,t_{0})=c_{1}(t-t_{0})^{1/3}+c_{2}\text{. }
\end{equation*}

We tested other powers except for $1/3$, as it provokes interesting discussions among more skeptical readers, even though it does not visually fit well. The aim here is to demonstrate that even when we encounter an inflection point where the second derivative is not defined, as suggested by the plots presented in Figure \ref{Fig:corroboration_inflection_point}, we encountered no issues. For the best fit, we obtained $c_{1}\approx -36$ and $c_{2}\approx 29$ for $t_{0}=1$ (fixed). Despite $\frac{d^{2}var(t,1)}{dt^{2}}=\frac{5}{9}c_{1}(t-1)^{-5/3}$ and $t_{0}=1$, ensuring an inflection point even when:
\begin{equation*}
\lim_{t\rightarrow 0^{+}}\frac{d^{2}var(t)}{dt^{2}}=+\infty \text{ and }%
\lim_{t\rightarrow 0^{-}}\frac{d^{2}var(t)}{dt^{2}}=-\infty
\end{equation*}
This scenario is a basic calculus concept, yet it is always worth noting.


\begin{thebibliography}{99}

\bibitem{Stanley} {H. Eugene Stanley, Introduction to Phase Transitions and Critical Phenomena, Oxford Science Publications, (1987)}

\bibitem{Bouchaud} {J-P Bouchaud, M. Potters, Theory of Financial Risk and Derivative Pricing, Cambridge University Press (2003)}

\bibitem{Castellano} {C. Castellano, S. Fortunato, V. Loreto, Statistical physics of social dynamics, Rev. Mod. Phys \textbf{81}, 591 (2009)}

\bibitem{Szabo} {G. Szabo, G. Fath, Evolutionary games on graphs, Phys. Rep. \textbf{446}, 97 (2007)}

\bibitem{Barabasi} {A.-L. Barabasi, Network Science, Cambridge University Press (2016)}

\bibitem{Janssen} {H. K. Janssen, B. Schaub, B. Schmittmann, New universal short-time scaling behaviour of critical relaxation processes, Z. Phys. B: Condens. Matter \textbf{73}, 539 (1989)}

\bibitem{Janssen2} {H. K. Janssen, K. Oerding, Non-equilibrium relaxation at a tricritical point, J. Phys. A: Math. Gen. \textbf{27}, 715 (1994)}

\bibitem{Henkel} M. Henkel, M. Pleimling, Non-equilibrium Phase Transitions, Vol. 2: Ageing and Dynamical Scaling far from Equilibrium, Springer, Dordrecht (2010)

\bibitem{Zhengprimordial} B. Zheng, Monte Carlo simulations of short-time critical dynamics, Int. J. Mod. Phys. B \textbf{12}, 1419 (1998)

\bibitem{Huse} D. A. Huse, Remanent magnetization decay at the spin-glass critical point: A new dynamic critical exponent for nonequilibrium autocorrelations, Phys. Rev. B \textbf{40}, 304 (1989)

\bibitem{Albano} E. V. Albano, M. A. Bab, G. Baglietto, R. A. Borzi, T. S. Grigera, E. S. Loscar, D. E. Rodriguez, M. L. R. Puzzo, G. P. Saracco, Study of phase transitions from short-time non-equilibrium behaviour, Rep. Prog. Phys. \textbf{74}, 026501 (2011)

\bibitem{Grinstein} G. Grinstein, C. Jayaprakash, Y. He, Statistical mechanics of probabilistic cellular automata, Phys. Rev. Lett. \textbf{55}, 2527 (1985)

\bibitem{TomePRE1998} T. Tome, M. J. de Oliveira, Short-time dynamics of critical nonequilibrium spin models, Phys. Rev. E \textbf{58}, 4242 (1998)

\bibitem{TaniaMario2014} T. Tome, M. de Oliveira, Stochastic Dynamics and Irreversibility, Springer (2014)

\bibitem{Janssen3} H. K. Janssen, On the nonequilibrium phase transition in reaction-diffusion systems with an absorbing stationary state, Z. Phys. B \textbf{42}, 151 (1981)

\bibitem{Grassberger} P. Grassberger, The critical behaviour of two-dimensional self-avoiding random walks, Z. Phys. B \textbf{48}, 255 (1982)

\bibitem{Dickman} J. Marro, R. Dickman, Nonequilibrium Phase Transitions, Cambridge (1999)

\bibitem{Hinrichsen2000} H. Hinrichsen, Non-equilibrium critical phenomena and phase transitions into absorbing states, Adv. Phys. \textbf{49}, 815 (2000)

\bibitem{SilvaDickman} R. da Silva, R. Dickman, J. R. Drugowich de Felicio, Critical behavior of nonequilibrium models in short-time Monte Carlo simulations, Phys. Rev. E \textbf{70}, 067701 (2004)

\bibitem{Janssen1994} H. K. Janssen, K. Oerding, Non-equilibrium relaxation at a tricritical point, J. Phys. A: Math. Gen. \textbf{27}, 715 (1994)

\bibitem{SilvaBC3D2022} R. da Silva, Numerical evidence of Janssen-Oerding's prediction in a three-dimensional spin model far from equilibrium, Phys. Rev. E \textbf{105}, 034114 (2022)

\bibitem{SilvaBJPMF} R. da Silva, Exploring the Similarities Between Mean-field and Short-range Relaxation Dynamics of Spin Models, Braz. J. Phys. \textbf{52}, 128 (2022)

\bibitem{Silva2002} R. da Silva, N. A. Alves, and J. R. Drugowich de Felicio, Mixed initial conditions to estimate the dynamic critical exponent in short-time Monte Carlo simulation, Phys. Rev. E \textbf{66}, 026130 (2002)

\bibitem{Silva2013} R. da Silva, H. A. Fernandes, J. R. Drugowich de Felicio, W. Figueiredo, Novel considerations about the non-equilibrium regime of the tricritical point in a metamagnetic model: Localization and tricritical exponents, Comp. Phys. Comm. \textbf{184}, 2371 (2013)

\bibitem{Wigner} E. P. Wigner, Characteristic Vectors of Bordered Matrices 
with Infinite Dimensions I, Ann. Math. \textbf{62}, 548--564 (1955)

\bibitem{Wignerb} E. P. Wigner, Characteristic Vectors of Bordered Matrices 
with Infinite Dimensions II, Ann. Math. \textbf{65}, 203--207 (1957)

\bibitem{Wigner2} E. P. Wigner, On the Distribution of the Roots of Certain Symmetric Matrices, Ann. Math. \textbf{67}, 325--327 (1958)

\bibitem{Mehta} M. L. Mehta, Random Matrices, Academic Press, Boston (1991)

\bibitem{Dyson} F.J. Dyson, Statistical Theory of the Energy Levels of Complex Sytems I, J. Math. Phys. \textbf{3}, 140--156 (1962)

\bibitem{Dyson2} F.J. Dyson, Statistical Theory of the Energy Levels of Complex Sytems II, J. Math. Phys. \textbf{3}, 157--165 (1962)

\bibitem{Dyson3} F.J. Dyson, Statistical Theory of the Energy Levels of Complex Sytems III, J. Math. Phys. \textbf{3}, 166--175 (1962)

\bibitem{Wishart} J. Wishart, The generalised product moment distribution in samples from a normal multivariate population, Biometrika \textbf{20A}, 32 (1928)

\bibitem{Livan} G. Livan, M. Novaes, P. Vivo, Introduction to Random Matrices, Theory and Practice, Springer (2018)

\bibitem{Vinayak2014} T. Vinayak, T. Prosen, B. Buca, T. H. Seligman, Spectral analysis of finite-time correlation matrices near equilibrium phase transitions, EPL \textbf{108}, 20006 (2014)

\bibitem{Biswas2017} S. Biswas, F. Leyvraz, P.M. Castillero, T.H. Seligman, Rich structure in the correlation matrix spectra in non-equilibrium steady states, Sci. Rep. \textbf{7} 40506 (2017)

\bibitem{RMT2023} R. da Silva, Random matrices theory elucidates the nonequilibrium critical phenomena, Int. J. Mod. Phys. C, \textbf{2350061} 1 (2023)

\bibitem{RMT-2} R. da Silva, H. C. M. Fernandes, E. Venites Filho, S. D. Prado, J. R. Drugowich de Felicio, Mean-field criticality explained by random matrices theory, Braz. J. Phys. \textbf{53}, \ 80 (2023)

\bibitem{RMTMaillard} J.-Ch. Angles d'Auriac, J.-M. Maillard, Random matrix theory in lattice statistical mechanics, Physica A \textbf{321}, 325 (2003)

\bibitem{Griffiths} R. B. Griffiths, Thermodynamics Near the Two-Fluid Critical Mixing Point in He3-He4, Phys. Rev. Lett. \textbf{24}, 715 (1970)

\bibitem{Blume1971} M. Blume, V. J. Emery, R. B. Griffiths, Ising Model for the $\lambda$ Transition and Phase Separation in He3-He4 Mixtures, Phys. Rev. A  \textbf{4}, 1071 (1971)

\bibitem{Blume} M. Blume, Theory of the First-Order Magnetic Phase Change in UO2, Phys. Rev. \textbf{141}, 517 (1966)

\bibitem{Capel} H. Capel, On the possibility transitions of first-order in Ising systems with zero-field phase of triplet ions splitting, Physica \textbf{32}, 966 (1966)

\bibitem{Butera} P. Butera, M. Pernici, The Blume-Capel model for spins S=1 and 3/2 in dimensions d=2 and 3, Physica A \textbf{507}, 22--66 (2018)

\bibitem{Beale} P.D. Beale, Finite-size scaling study of the two-dimensional Blume-Capel model, Phys. Rev. B \textbf{33}, 1717--1720 (1986)

\bibitem{Deserno} M. Deserno, Tricriticality and the Blume-Capel model: A Monte Carlo study within the microcanonical ensemble, Phys. Rev. E \textbf{56}, 5204--5210 (1997)

\bibitem{Plascak} J.C. Xavier, F. C. Alcaraz, D. Pena Lara, J. A. Plascak, Critical behavior of the spin-3/2 Blume-Capel model in two dimensions, Phys. Rev. B \textbf{57}, 11575 (1998)

\bibitem{Hasenbusch} M. Hasenbusch, Dynamic critical exponent $z$ of the three-dimensional Ising universality class: Monte Carlo simulations of the improved Blume-Capel model, Phys. Rev. E. \textbf{101}, 022126 (2020)

\bibitem{Marcenko} V. A. Marcenko, L. A. Pastur, Distribution of Eigenvalues for Some Sets of Random Matrices, Math. USSR Sb. \textbf{1} 457 (1967)

\bibitem{Baik2005} J. Baik, G. B. Arous, S. Pechet, Phase transition of the largest eigenvalue for nonnull complex sample covariance matrices, Ann. Probab. \textbf{33}, 1643 (2005)

\bibitem{Benaych2011} F. Benaych-Georges, A. Guionnet, M. Maida, Fluctuations of the extreme eigenvalues of finite rank deformations of random matrices, Electron. J. Probab. \textbf{16}, 1621 (2011)

\bibitem{SilvaDynamics} R. da Silva, S. D. Prado, Exploring Transition from Stability to Chaos through Random Matrices, Dynamics \textbf{3}, 777-792 (2023)

\bibitem{PRLnew} R. da Silva, T. Tome, M. J. de Oliveira, Numerical exploration of the aging effects in spin systems, Phys. Lett. A \textbf{489}, 129148 (2023)


\end{thebibliography}
\end{document}